\documentclass[11pt]{article}

\usepackage{amsmath,amssymb,amsthm}
\usepackage[margin=1in]{geometry}
\usepackage[numbers,sort&compress]{natbib}
\usepackage{hyperref}
\usepackage{enumitem}
\usepackage{microtype}
\usepackage{tikz}
\usetikzlibrary{positioning,calc,fit,backgrounds,shapes.geometric,shapes.misc,%
                decorations.pathreplacing}

\definecolor{cInk}{HTML}{1F2A37}
\definecolor{cPaper}{HTML}{FBFAF7}
\definecolor{cRule}{HTML}{D6D1C7}
\definecolor{cMute}{HTML}{9B958A}
\definecolor{cGA}{HTML}{35547F}
\definecolor{cGB}{HTML}{A97C42}
\definecolor{cGC}{HTML}{4E7A5E}
\definecolor{cGD}{HTML}{6E5C88}
\definecolor{cAcc}{HTML}{9C3B33}

\definecolor{cCite}{HTML}{2F6B4F}
\definecolor{cLink}{HTML}{35547F}
\definecolor{cUrl}{HTML}{9C3B33}
\hypersetup{
  colorlinks = true,
  citecolor  = cCite,
  linkcolor  = cLink,
  urlcolor   = cUrl,
  linktoc    = all,
  pdfborder  = {0 0 0}
}

\usepackage[framemethod=TikZ]{mdframed}
\newmdenv[
  roundcorner        = 7pt,
  backgroundcolor    = cInk!5,
  linecolor          = cInk!30,
  linewidth          = 0.7pt,
  innerleftmargin    = 11pt,
  innerrightmargin   = 11pt,
  innertopmargin     = 9pt,
  innerbottommargin  = 9pt,
  skipabove          = \medskipamount,
  skipbelow          = \medskipamount
]{algbox}

\tikzset{
  >=stealth,
  gpA/.style={circle,draw=cGA,fill=cGA!30,line width=0.6pt,
              minimum size=5.6pt,inner sep=0pt},
  gpB/.style={rectangle,draw=cGB,fill=cGB!35,line width=0.6pt,
              minimum size=5.4pt,inner sep=0pt},
  gpC/.style={regular polygon,regular polygon sides=3,draw=cGC,fill=cGC!30,
              line width=0.6pt,minimum size=7pt,inner sep=0pt},
  gpD/.style={diamond,draw=cGD,fill=cGD!35,line width=0.6pt,
              minimum size=6.6pt,inner sep=0pt},
  tnode/.style={circle,draw=cGA!75,fill=cGA!9,line width=0.7pt,
                minimum size=17pt,inner sep=1pt,font=\small,text=cInk},
  vnode/.style={circle,draw=cGC!75,fill=cGC!9,line width=0.7pt,
                minimum size=22pt,inner sep=1pt,font=\small,text=cInk},
  panel/.style={rounded corners=2.5pt,draw=cRule,fill=cPaper,line width=0.6pt},
  card/.style={rounded corners=2.5pt,draw=cRule,fill=cPaper,line width=0.6pt},
  lnk/.style={cMute!75,line width=0.5pt},
  lab/.style={font=\footnotesize,text=cInk},
  plabel/.style={font=\small\itshape,text=cInk!85},
  note/.style={font=\scriptsize,text=cInk!72},
}

\newtheorem{theorem}{Theorem}
\newtheorem{lemma}[theorem]{Lemma}
\newtheorem{proposition}[theorem]{Proposition}
\newtheorem{corollary}[theorem]{Corollary}
\newtheorem{observation}[theorem]{Observation}
\theoremstyle{definition}
\newtheorem{definition}[theorem]{Definition}
\newtheorem{remark}[theorem]{Remark}

\newcommand{\OPT}{\mathrm{OPT}}
\newcommand{\cost}{\operatorname{cost}}
\newcommand{\E}{\mathbb{E}}
\newcommand{\R}{\mathbb{R}}
\newcommand{\Z}{\mathbb{Z}}
\newcommand{\eps}{\varepsilon}
\newcommand{\rt}{\mathrm{root}}
\DeclareMathOperator{\dist}{dist}

\definecolor{cReply}{HTML}{0F6E6E}

\newif\ifshowrev
\showrevfalse       
\newif\ifdelrevstrike \delrevstriketrue
\newif\ifhasulem
\IfFileExists{ulem.sty}{\hasulemtrue}{\hasulemfalse}
\ifdelrevstrike\else\hasulemfalse\fi
\ifhasulem
  \usepackage[normalem]{ulem}
  \newcommand{\delmark}[1]{\sout{#1}}
\else
  \newcommand{\delmark}[1]{[removed: #1]}
\fi
\definecolor{cAdd}{HTML}{1345C4}   
\definecolor{cDel}{HTML}{C0181A}   
\definecolor{cNote}{HTML}{15803D}  

\makeatletter
\ifshowrev
  \newcommand{\rev}[1]{{\color{cAdd}#1}}
  \newcommand{\delrev}[1]{\leavevmode{\color{cDel}\delmark{#1}}}
  \newcommand{\delnote}[1]{\leavevmode{\color{cDel}\textsf{\small[#1]}}}
  \newcommand{\ainote}[1]{\leavevmode{\color{cNote}\textsf{\small[Note to Ali: #1]}}}
\else
  \newcommand{\rev}[1]{#1}
  \newcommand{\delrev}[1]{\@bsphack\@esphack}
  \newcommand{\delnote}[1]{\@bsphack\@esphack}
  \newcommand{\ainote}[1]{\@bsphack\@esphack}
\fi
\makeatother

\title{Tight Approximations for Pairwise Fair Clustering via Scalar Repair on Tree Embeddings}
\author{Ali Vakilian\thanks{Rutgers University. Email: \texttt{a.vakilian@rutgers.edu}. The author was supported in part by NSF CAREER award CAIF-2638918.}}
\date{}

\begin{document}

\maketitle

\begin{abstract}
In pairwise fair clustering, the input points are partitioned into $\ell$ protected
groups, and a clustering is called fair only if, inside every cluster, no group dominates any other group by more than a factor $t$: the number of points of any group is at most $t$ times the number of points of any other.
We study algorithms that never violate this constraint and approximate the clustering cost.
For the $k$-center objective such algorithms achieve a constant approximation ratio in polynomial time, while for $k$-median the best ratio achieved in polynomial time was $O(k^{2}\ell t)$. For $k$-means with more than two groups, polynomial-time algorithms were known only for perfect balance ($t=1$); beyond that, known algorithms violate the constraint or run in exponential time.

We give a single randomized polynomial-time algorithm that improves both cases. For pairwise fair $k$-median it
returns a clustering that is always exactly fair and whose expected cost is within a factor $O(\log k)$ of the cheapest fair clustering, removing all dependence on the number of groups $\ell$ and on the balance parameter $t$. This is best possible up to
constants short of a breakthrough: an $o(\log k)$ bound would give an $o(\log k)$
approximation for uniform capacitated $k$-median, where $O(\log k)$ has stood for decades.

We then extend the algorithm to the general $(k,z)$-clustering objective and obtain
approximation ratio $2^{O(z)}k^{z-1}\log k$, again for the expected cost and with exact fairness.
Besides $k$-median, which is the case $z=1$, a second use case of this result is
$z=2$: it gives an $O(k\log k)$-approximation for pairwise fair $k$-means with exact fairness, the first such guarantee for general $t$. All results hold for every real $t\ge 1$, including non-integral thresholds such as the $t=5/4$ of the ``four-fifths rule''.

The algorithm relies on a structural observation: on integral cluster loads, pairwise fairness is equivalent to the existence of a single nonnegative integer base $h$ with $h \le (\text{load of each group}) \le t\cdot h$, and these bases add along a hierarchical decomposition. This collapses an apparently $\ell$-dimensional feasibility condition to one scalar, so that a dynamic program on a tree embedding of the centers can carry one integer per subtree instead of $\ell$ group counts. The program minimizes a one-dimensional lower bound on the cost of repairing an unfair assignment, and a top-down pass turns its solution into an exactly fair assignment while losing only a constant factor $4$.
\end{abstract}

\section{Introduction}
\label{sec:intro}

Clustering is rarely the end of a pipeline. A partition of customers into segments
decides who receives which offer; a partition of a student cohort into sections
decides who studies with whom; a partition of a service region into districts decides who is served by which facility. When a cluster is the unit at which a decision is applied, the demographic composition of that cluster is inherited by everyone in it. An algorithm that is indifferent to composition will return clusters that are nearly homogeneous, because homogeneous clusters are often exactly the ones that minimize distance-based cost. The result is a partition in which some groups are concentrated in a few clusters and absent from the rest, so that any downstream
treatment applied per cluster falls unevenly across groups
\citep{BarocasS16,DworkHPRZ12}.

The line of work started by \citet{ChierichettiKLV17} addresses this by constraining
the composition of each cluster directly. Their original setting has two groups,
say red and blue, and a balance parameter $t \ge 1$: a clustering is \emph{fair} if in
every cluster the number of red points is at most $t$ times the number of blue points
and vice versa. With $t=1$ every cluster is perfectly balanced; as $t$ grows the
constraint relaxes, and for $t \ge n$ it disappears. The parameter is a single
readable knob that interpolates between full balance and no constraint at all.

\medskip
Real data has more than two groups. The generalization we study, following \citet*{BandyapadhyayFS24} and \citet{BandyapadhyayCFJMV26}, applies the same ratio
test to every ordered pair of groups.

\begin{definition}[Pairwise fair clustering]
\label{def:pfc}
The input is a metric space $(\Omega,d)$, a set $P \subseteq \Omega$ of $n$ points
partitioned into $\ell$ disjoint groups $P_1,\dots,P_\ell$, an integer $k \ge 1$, and
a balance parameter $t \ge 1$. A clustering $\mathcal{C}$ of $P$ is
\emph{pairwise fair} (or \emph{$t$-fair}) if for every cluster  $C \in \mathcal{C}$ and all $i,j \in [\ell]$,
\begin{align*}
    |C \cap P_i| \;\le\; t\,|C \cap P_j|.
\end{align*}

The goal is to open at most $k$ centers and to assign every point to an open center so
that the induced clustering is pairwise fair and a given cost is minimized.
\end{definition}

Because the condition is required for every ordered pair $(i,j)$, it is equivalent to
the compact statement
$\max_{a} |C \cap P_a| \le t \min_{a} |C \cap P_a|$,
and in particular a nonempty fair cluster must contain at least one point of every
group. We write $\cost_z(\phi) = \sum_{p \in P} d(p,\phi(p))^z$ for an assignment
$\phi$; the exponent $z=1$ gives $k$-median, $z=2$ gives $k$-means, and general
constant $z \ge 1$ gives $(k,z)$-clustering.

Two features make this notion appealing in practice. First, it is parameterized by one
number rather than by $2\ell$ numbers. The competing and also well-studied notion of
\emph{representational} fairness \citep{BeraCFN19,BerceaGKKRSS19} asks that the
\emph{fraction} of each group $i$ in each cluster lie in an interval
$[\alpha_i,\beta_i]$, which requires the analyst to choose and defend $2\ell$
thresholds and makes it easy to write down an infeasible instance by accident.
Pairwise fairness instead adapts automatically: it is feasible exactly when the whole
data set passes the same ratio test (Section~\ref{sec:prelim}), and it never asks the
analyst to know the group proportions in advance.
Second, a ratio between group counts is the form that composition requirements
already take in practice. The best-known instance is the ``four-fifths rule'' of the
Uniform Guidelines on Employee Selection Procedures \citep{UGESP78}, under which a
selection rate below $4/5$ of the highest group's rate is treated as evidence of adverse impact; it is a ratio test with $t = 5/4$, stated for rates rather than counts, and it is the standard operational test in the disparate-impact literature \citep{FeldmanFMSV15}.

\paragraph{Exact fairness.} Throughout, an algorithm is \emph{exactly fair} if every cluster of its output satisfies Definition~\ref{def:pfc} for the given $t$, with no additive or multiplicative slack. (``Exact'' refers to the absence of violations, not to the perfectly balanced case $t=1$.) Much of the literature instead gives \emph{bicriteria} guarantees, which allow each group count in each cluster to be off by a small additive term \citep{BerceaGKKRSS19,BeraCFN19}. Additive slack is harmless for large clusters but weak precisely where composition matters most: in a small cluster, a slack of a single point can remove a group altogether. 
We therefore ask, as \citet{BandyapadhyayCFJ25} and \citet{BandyapadhyayCFJMV26} do, for polynomial-time algorithms with no violation at all. What is approximated is only the cost.

\medskip
Exact pairwise fairness has been difficult.
For two groups it coincides with representational fairness with thresholds $1/(t+1)$ and $t/(t+1)$, and the fairlet technique of \citet{ChierichettiKLV17} gives an $O(t)$-approximation for $k$-median, which is a constant only when $t$ is.
For $\ell > 2$ the picture until recently was:

\begin{itemize}[leftmargin=*]
\item \textbf{Parameterized time.} \citet*{BandyapadhyayFS24} give a
constant-factor approximation for pairwise fair $k$-median and $k$-means, exactly fair,
in time $(k\ell)^{O(k)}n^{O(1)}$. \citet*{DaiMV22} give an $O(\log k)$ approximation for
the representational variant in time $n^{O(\ell)}$. Neither is polynomial.
\item \textbf{$k$-center.} \citet{BandyapadhyayCFJ25} give the first polynomial-time
$O(1)$ approximation for pairwise fair $k$-center with arbitrary $\ell$ and integer
$t$, via a tree-based rounding of a linear program. They also show that the same
relaxation has an $\Omega(k)$ integrality gap when $t$ is not an integer, even for two
groups.
\item \textbf{$k$-median.} \citet{BandyapadhyayCFJMV26} give the first polynomial-time exactly fair approximation, with ratio $O(k^{2}\ell t)$.
\item \textbf{$k$-means.} For perfect balance ($t=1$), where every cluster holds the same number of points of each group, \citet{BohmFLS20} reduce the problem to the unconstrained one at the loss of a constant factor, for $k$-median, $k$-center and $k$-means. For $\ell>2$ and $t>1$ nothing polynomial-time and exactly fair was known.
\end{itemize}

The $k$-median bound $O(k^{2}\ell t)$ is unsatisfying in a specific way: it degrades
with quantities that have nothing to do with the geometry.
It grows linearly with the number of groups $\ell$ and with the balance parameter $t$, even though raising $t$ only relaxes the constraint and can only make the problem easier. The dependence on $\ell$ matters in practice: when the groups are the intersections of several protected attributes, every additional binary attribute doubles $\ell$.
The reason, we show in Section~\ref{sec:limits}, is not an artifact of the analysis but
of the relaxation: the natural assignment linear program has an integrality gap of
$\Omega(\ell t)$ even on an instance with two locations on a line.

For $k$-means the obstacle is specific to the route through trees. Squared distances violate the triangle inequality, but only mildly as long as it is applied a bounded number of times, since $(a+b)^{2}\le 2a^{2}+2b^{2}$. This is why fairlet and LP-rounding arguments extend to $k$-means when small violations or perfect balance are allowed \citep{SchmidtSS19,BeraCFN19,BerceaGKKRSS19,BohmFLS20}. Our algorithm, like that of \citet*{DaiMV22}, instead embeds the centers into a tree, and an embedding compares a distance with a path of many hops. There the failure is not mild:
on the points $\{0,1,\dots,q\}$ of the real line, the squared distance between
the endpoints is $q^{2}$ while the sum of the $q$ unit steps is only $q$.

\subsection{Our results}

We give one algorithm and one analysis that covers all of these objectives, with a
guarantee that depends only on $k$.

\begin{theorem}[$k$-median; see Theorem~\ref{thm:median}]
\label{thm:intro-median}
There is a randomized polynomial-time algorithm for pairwise fair
$k$-median that opens at most $k$ centers, satisfies every pairwise fairness
constraint exactly, and outputs a solution of expected cost at most $O(\log k)\cdot \OPT$.
\end{theorem}

Our algorithm calls a polynomial-time algorithm for \emph{unconstrained} $k$-median as a subroutine, and if this subroutine has approximation ratio $\alpha$, then Theorem~\ref{thm:median} gives an $O(\alpha\log k)$-approximation. For $k$-median, several constant-factor approximations are known, yielding the stated $O(\log k)$.
For a general exponent $z\ge1$ the same algorithm applies, and two things change in the bound. First, the unconstrained subroutine has ratio $\alpha_z=2^{O(z)}$ (Section~\ref{sec:related}). Second, $d^{z}$ is not a metric when $z>1$, and before the centers can be embedded into a tree it must be replaced by one; on $k$ centers this costs a factor $k^{z-1}$ (Lemma~\ref{lem:metric}). These two factors are the only difference between Theorem~\ref{thm:intro-median} and the next theorem.

\begin{theorem}[$(k,z)$-clustering; see Theorem~\ref{thm:kz}]
\label{thm:intro-kz}
Let $z \ge 1$.
There is a randomized polynomial-time algorithm for
pairwise fair $(k,z)$-clustering that is exactly fair and has expected cost at most $2^{O(z)}k^{z-1}\log k \cdot \OPT$.
\end{theorem}

\begin{corollary}[$k$-means]
\label{cor:intro-means}
Pairwise fair $k$-means admits a randomized polynomial-time exactly fair approximation
with expected cost at most $O(k\log k)\cdot\OPT$.
\end{corollary}

All results hold in an arbitrary metric space.
Only the cost is random: the output is exactly fair for every outcome of the random choices. The bound on the expected cost becomes a high-probability bound through independent repetition.
Moreover, Theorems~\ref{thm:intro-median} and \ref{thm:intro-kz} hold for every real
$t \ge 1$, not only integral $t$.
This is worth stating separately because integrality of $t$ is not a technicality elsewhere: for $k$-center, the relaxation of \citet{BandyapadhyayCFJ25} has
an $\Omega(k)$ integrality gap when $t$ is fractional. Our dynamic program never forms that relaxation, so it is insensitive to the arithmetic of $t$.

\smallskip
By the reduction of \citet{BandyapadhyayCFJMV26}, a polynomial-time $\alpha$-approximation for pairwise fair $k$-median with only two groups implies a
$(1+\eps)\alpha$ approximation for soft uniform capacitated $k$-median with the same
$k$. For that problem the best known polynomial-time guarantee without any violation
has been $O(\log k)$, obtained by embedding into trees and solving the tree instance
by dynamic programming \citep{CharikarCGG98,Bartal96}, and whether a constant factor is
achievable is a well-known open question \citep{CohenAddadL19,Li17,CohenAddad20}.
Theorem~\ref{thm:intro-median} therefore hits exactly the same barrier.
We call the bound tight in this conditional sense, and we do not
claim an unconditional lower bound.

We complement the algorithms with a limitation, proved in Section~\ref{sec:limits}. It explains why a guarantee free of $\ell$ and $t$ could not come from the
previous framework.

\begin{proposition}[see Proposition~\ref{prop:lpgap}]
\label{prop:intro-lpgap}
For every $\ell \ge 2$, every real $t \ge 2$ and every $z \ge 1$, the natural fair
assignment relaxation has integrality gap at least $(\ell-1)t$, already on an instance
with two candidate centers at distance $1$ on a line.
\end{proposition}

We also note that the $k^{z-1}$ factor in Theorem~\ref{thm:intro-kz}  is simply what it costs to replace $d^{z}$ by a metric, and no sharper choice of metric would avoid it (Remark~\ref{rem:metrictight}).
So the $O(k\log k)$ of Corollary~\ref{cor:intro-means} is the best this route gives,
and improving it requires an argument that charges the transported mass rather than the worst distorted pair.

\subsection{Technical overview}
\label{sec:overview}

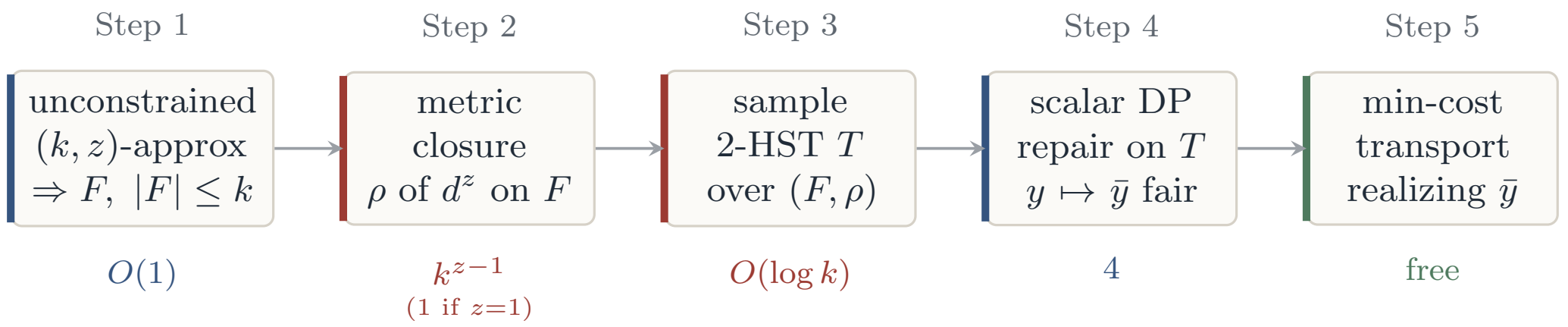
\begin{figure}[t]
\centering
\begin{tikzpicture}[node distance=6mm]
\tikzset{pbox/.style={card,align=center,font=\footnotesize,text=cInk,
                      inner sep=4pt,minimum height=13mm}}
\node[pbox,text width=20mm] (s1)
  {unconstrained\\ $(k,z)$-approx\\ $\Rightarrow F,\ |F|\le k$};
\node[pbox,text width=19mm,right=of s1] (s2)
  {metric closure\\ $\rho$ of $d^z$ on $F$};
\node[pbox,text width=19mm,right=of s2] (s3)
  {sample\\ $2$-HST $T$\\ over $(F,\rho)$};
\node[pbox,text width=19mm,right=of s3] (s4)
  {scalar DP\\ repair on $T$\\ $y\mapsto\bar y$ fair};
\node[pbox,text width=19mm,right=of s4] (s5)
  {min-cost\\ transport\\ realizing $\bar y$};
\foreach \n/\c in {s1/cGA,s2/cAcc,s3/cAcc,s4/cGA,s5/cGC}
  {\draw[\c,line width=1.9pt] ([yshift=1.2pt]\n.south west) -- ([yshift=-1.2pt]\n.north west);}
\foreach \a/\b in {s1/s2,s2/s3,s3/s4,s4/s5}
  {\draw[->,cInk!45,line width=0.7pt] (\a) -- (\b);}
\foreach \n/\s in {s1/1,s2/2,s3/3,s4/4,s5/5}
  {\node[note,above=1.1mm of \n] {Step \s};}
\node[note,below=1.4mm of s1,text=cGA]  {$O(1)$};
\node[note,below=1.4mm of s2,text=cAcc,align=center] {$k^{z-1}$\\[-2pt] {\tiny($1$ if $z{=}1$)}};
\node[note,below=1.4mm of s3,text=cAcc] {$O(\log k)$};
\node[note,below=1.4mm of s4,text=cGA]  {$4$};
\node[note,below=1.4mm of s5,text=cGC]  {free};
\end{tikzpicture}
\caption[Overview of the algorithm]{Overview of our algorithm and where its guarantee is spent. The dynamic program is the only step that touches the fairness constraint. The symbols are defined in the text below: $F$ is the set of centers, $\rho$ the metric of Step~2, $T$ the tree of Step~3, $y$ the group counts of the nearest-center assignment and $\bar y$ the repaired, fair counts.}
\label{fig:pipeline}
\end{figure}

The algorithm has five steps (Figure~\ref{fig:pipeline}). Steps~1--3 and~5 are assembled from known tools. Their role is to reduce the entire problem to a single combinatorial question on a tree whose leaves are at most $k$ centers, which we call \emph{repair}. Step~4 solves repair, and it is where our contribution lies. We describe the reduction briefly and Step~4 in detail. This overview is self-contained; Sections~\ref{sec:prelim}--\ref{sec:main} supply the proofs.

\paragraph{Steps 1--3: reducing to repair on a tree.}
\emph{Step 1 (centers).} Run any $\alpha_z$-approximation for the \emph{unconstrained} $(k,z)$-clustering problem, let $F$ be its set of centers, $m=|F|\le k$, and let $\sigma$ assign every point to its nearest center in $F$. Every fair solution is in particular an unconstrained one, so $\cost_z(\sigma)\le\alpha_z\OPT$. The assignment $\sigma$ is cheap but in general unfair. However, $F$ also supports a cheap \emph{fair} assignment $\gamma\colon P\to F$, of cost at most $2^{O(z)}(1+\alpha_z)\,\OPT$: send each cluster of an optimal fair solution, as a whole, to the center of $F$ nearest to its optimal center. The triangle inequality bounds the cost (Lemma~\ref{lem:projection}), and $\gamma$ is fair because a union of fair clusters is fair (Observation~\ref{obs:union}). This step is the same as in \citet*{DaiMV22} and \citet{BandyapadhyayCFJMV26}; we state it for general $z$. The algorithm cannot compute $\gamma$. It only uses that $\gamma$ exists.

From here on the centers are fixed and only the assignment changes. All that matters about an assignment is its \emph{load matrix}: $y_{fa}$ denotes the number of points of group $a$ assigned to center $f$. An assignment is fair exactly when every row of its load matrix is fair, that is, $\max_a y_{fa}\le t\min_a y_{fa}$ for every $f$. Let $y$ be the load matrix of $\sigma$. We want to make $\sigma$ fair by \emph{moving} points, that is, by reassigning some points from their current center $f$ to another center $g$. A moved point then pays, up to a constant, what it paid before plus $d(f,g)^{z}$.

\emph{Step 2 (a metric on the centers).} For $z=1$ the moving cost $d$ is a metric on $F$ and we set $\rho=d$. For $z>1$, $d^{z}$ is not a metric, and we replace it by the cost of the cheapest route from $f$ to $g$ through other centers: $\rho(f,g)$ is the minimum of $\sum_i d(v_i,v_{i+1})^{z}$ over all paths $f=v_0,v_1,\dots,v_r=g$ in $F$. This $\rho$ is a metric with $\rho\le d^{z}\le(m-1)^{z-1}\rho$, where $m=|F|\le k$ is the number of centers (Lemma~\ref{lem:metric}): a cheapest route can be taken to visit no center twice, so it has at most $m-1$ hops. The factor $(m-1)^{z-1}\le k^{z-1}$ is the entire difference between our bounds for $k$-median and for $k$-means.

\emph{Step 3 (a tree).} Using the embedding of \citet*{FakcharoenpholRT04} we sample a tree $T$ whose leaves are the centers. Its tree distance $\tau$ satisfies $\rho\le\tau$ always and $\E[\tau(f,g)]\le O(\log k)\,\rho(f,g)$; the logarithm is in $k$ and not in $n$ because only the $m\le k$ centers are embedded. The tree is a \emph{$2$-HST}: all leaves have the same depth, all edges entering the same level have the same weight, and the weights halve from each level to the next. For a node $u$ of $T$ we write $F_u$ for the set of centers below $u$, $w_u$ for the weight of the edge joining $u$ to its parent, and
\[
  y_a(u)\;=\;\sum_{f\in F_u}y_{fa}
\]
for the number of points of group $a$ currently assigned inside the subtree of $u$.

Moving a point from leaf $f$ to leaf $g$ costs $\tau(f,g)$, the total weight of the tree path between them. That path uses the edge above $u$ exactly when one of $f,g$ lies below $u$ and the other does not. So \emph{a point pays $w_u$ precisely when it is moved into or out of the subtree of $u$}, and we then say that it \emph{crosses} the edge above $u$. Now suppose the moves change the loads from $y$ to $\bar y$. The number of group-$a$ points inside the subtree of $u$ changes from $y_a(u)$ to $\bar y_a(u)$, so at least $|y_a(u)-\bar y_a(u)|$ points of group $a$ must cross the edge above $u$. Moving surpluses up the tree and deficits down achieves exactly this number at all edges simultaneously (Observation~\ref{obs:treeemd}). The cheapest way to change $y$ into $\bar y$ therefore costs
\[
  W_T(y,\bar y)\;=\;\sum_{a=1}^{\ell}\ \sum_{u}\,w_u\,\bigl|y_a(u)-\bar y_a(u)\bigr| ,
\]
where $u$ ranges over the nodes other than the root. This cost depends on the loads only through the group totals of the subtrees.

\emph{The repair problem.} Given the tree $T$ and the loads $y$ of $\sigma$, find a fair load matrix $\bar y$ with the same group totals as $y$ for which $W_T(y,\bar y)$ is small. A good target exists, namely the load matrix $y^\gamma$ of $\gamma$, but we cannot see it. The guarantee we need is therefore one against \emph{every} fair target. 

\paragraph{Step 4: repair with one integer per subtree.}
Even on a tree we do not know how to solve repair optimally in polynomial time. The obvious dynamic program records, for each subtree, how many points of each group it ends up holding. That is a vector of $\ell$ numbers, with $n^{\Theta(\ell)}$ possible values. This is the source of the $n^{O(\ell)}$ running time of \citet*{DaiMV22}, whose algorithm has the same architecture as ours (fixed centers, tree embedding, dynamic program) but addresses a different constraint, representational fairness with $2\ell$ thresholds. The comparison we draw with their work is one of technique and not of results: we do not improve their theorem, and we do not know whether their constraint admits a summary like the one below. Linear programming does not help either, since the natural relaxation has integrality gap $\Omega(\ell t)$ (Proposition~\ref{prop:intro-lpgap}). In capacitated $k$-median, by contrast, a dynamic program on a tree needs \emph{one} number per subtree, the number of clients it serves. Pairwise fairness constrains $\ell$ coupled counts per center, and it is not obvious which single number could play that role. Our answer is the base.

\emph{Bases and windows.} Assume for now that $t$ is an integer. For an integer $b\ge0$ call
\[
  I(b)\;=\;\{b,\ b+1,\ \dots,\ tb\}
\]
the \emph{window} of $b$; note that $I(0)=\{0\}$. A row of loads $(y_{f1},\dots,y_{f\ell})$ is fair if and only if all its entries lie in one common window: if the row is fair, take $b=\min_a y_{fa}$; conversely, if all entries lie in $I(b)$ then the largest is at most $tb$ and the smallest at least $b$. We turn this around and treat the integer as a \emph{decision}. A \emph{base vector} $h=(h_f)_{f\in F}$ assigns to each center a nonnegative integer $h_f$, its \emph{base}, and a load matrix $y'$ is \emph{compatible} with $h$ if $y'_{fa}\in I(h_f)$ for all $f$ and $a$. In words, the base is a quota: center $f$ must receive at least $h_f$ and at most $t\,h_f$ points of every group, and a center with base $0$ receives nothing, that is, it is closed. Every compatible load matrix is fair, and every fair load matrix is compatible with the vector of its row minima. So instead of searching for fair loads directly, we may search for a base vector and then for loads compatible with it. Lower and upper bounds of this kind on the group counts at a center are not new; they are also used by \citet{BandyapadhyayCFJMV26}. What is new is how they behave on a tree, and what we do with that.

\emph{Windows add.} Define the base of a node as the total base of the centers below it, $h(u)=\sum_{f\in F_u}h_f$. The group-$a$ total of a subtree is the sum of the group-$a$ loads of its leaves, so the totals that a subtree can hold are the sums of one number from each leaf's window. A window is a set of consecutive integers whose two ends, $b$ and $tb$, are proportional to $b$. Such sums therefore fill exactly the window of the sum (Lemma~\ref{lem:base}):
\begin{quote}
if $x_i\in I(b_i)$ for $i=1,\dots,s$, then $\sum_i x_i\in I\bigl(\sum_i b_i\bigr)$, and every element of $I\bigl(\sum_i b_i\bigr)$ can be written in this way.
\end{quote}
Hence, whatever loads compatible with $h$ the leaves carry, each group total of the subtree of $u$ lies in $I(h(u))$, and every value in $I(h(u))$ can be realized. What a subtree can hold is described by one integer, regardless of the number of groups and of how the base is spread among its leaves.

\emph{A lower bound that needs only this integer.} Suppose the subtree of $u$ has base $b$. Whatever compatible loads we end with, its group-$a$ total must land in $I(b)$, while currently it is $y_a(u)$. So at least
\[
  \dist\bigl(y_a(u),I(b)\bigr)\;=\;\max\bigl\{0,\ b-y_a(u),\ y_a(u)-tb\bigr\}
\]
points of group $a$ must cross the edge above $u$: inward if the subtree is short of group $a$, outward if it has too many. Summing over the groups gives the \emph{forced traffic} $\delta_u(b)=\sum_a\dist(y_a(u),I(b))$, and pricing it at every edge gives the \emph{surrogate cost} of a base vector,
\[
  \Phi_T(h)\;=\;\sum_{u}w_u\,\delta_u\bigl(h(u)\bigr).
\]
By construction $\Phi_T(h)\le W_T(y,y')$ for every load matrix $y'$ that is compatible with $h$ and has the group totals of $y$: term by term, $|y_a(u)-y'_a(u)|\ge\dist(y_a(u),I(h(u)))$ because $y'_a(u)$ lies in $I(h(u))$. The surrogate is only a lower bound, since it charges every edge the least traffic that this edge must carry, and no single repair need attain all these minima at once. But it depends on each subtree only through the integer $h(u)$.

\emph{Pass 1 (bottom-up): choose the bases.} We minimize $\Phi_T$ by dynamic programming. Let $V_u(b)$ be the smallest possible contribution of the subtree of $u$ to $\Phi_T$ when its base is $b$. A leaf has no choice. For an internal node with children $u_1,\dots,u_s$, the base $b$ is split among the children in the best way:
\[
  V_u(b)\;=\;w_u\,\delta_u(b)\;+\;\min_{b_1+\dots+b_s=b}\ \sum_{i=1}^{s}V_{u_i}(b_i).
\]
At the root the window must contain the actual group totals $N_a$, that is, $\lceil\max_aN_a/t\rceil\le b\le\min_aN_a$; such a $b$ exists exactly when the instance is feasible. Every table has at most $n+1$ entries, and the program runs in time $O(|T|(n^{2}+n\ell))$, polynomial in $\ell$. Let $h$ be the base vector it returns.

\emph{Pass 2 (top-down): turn bases into loads.} Handle each group $a$ separately. The root is given the target $N_a$, which lies in its window. A node that has received a target in its window divides it among its children. Each child is first offered the value of its own window closest to the count it currently holds. If the offers add up to less than the parent's target, the missing amount is handed out to the children one after another, each child's offer being raised as far as its window allows, until nothing is missing. If they add up to more, the excess is taken back in the same way, by lowering offers. So at any one node the offers are either only raised or only lowered. This is always possible because the children's windows add up to the parent's. The targets that reach the leaves form $\bar y$. They lie in the leaves' windows, so $\bar y$ is compatible with $h$ and hence exactly fair.

\emph{Why this costs at most $4\,\Phi_T(h)$.} Call the \emph{error} of a node the difference between its target and its current count, $|\bar y_a(u)-y_a(u)|$; the cost $W_T(y,\bar y)$ is the sum of $w_u$ times the errors, over nodes and groups. The offers to the children differ from the children's counts by exactly their forced traffic. The correction, being in one direction, moves the offers by a total equal to the gap between the parent's target and the sum of the offers, and this gap is at most the parent's error plus, once more, the children's forced traffic. So
\[
  \text{(total error of the children)}\;\le\;\text{(error of the parent)}\;+\;2\cdot\text{(forced traffic of the children)}.
\]
In a $2$-HST the edge above a child weighs half as much as the edge above its parent. Multiplying by the weights and summing over the tree, the cost $W=W_T(y,\bar y)$ satisfies $W\le\tfrac12W+2\,\Phi_T(h)$, that is, $W\le4\,\Phi_T(h)$.

\emph{The guarantee.} Let $y^\star$ be any fair load matrix with the same group totals as $y$, and let $h^\star$ be the vector of its row minima. Then $y^\star$ is compatible with $h^\star$, and $h^\star$ passes the test at the root, so it is one of the candidates of Pass~1. Therefore
\[
  W_T(y,\bar y)\;\le\;4\,\Phi_T(h)\;\le\;4\,\Phi_T(h^\star)\;\le\;4\,W_T(y,y^\star),
\]
the three inequalities being Pass~2, the optimality of $h$ in Pass~1, and the lower-bound property (Lemma~\ref{lem:repair}). In polynomial time, and knowing only $y$ and $T$, we obtain exactly fair loads whose repair cost is within a factor $4$ of that of the best fair target, in particular of $y^\gamma$.

\emph{Real $t$.} When $t$ is not an integer, the window of $b$ is $\{b,\dots,\lfloor tb\rfloor\}$. Its upper end is no longer proportional to $b$, so one integer does not determine what a subtree can hold. Two integers do: the sum of the bases and the sum of the upper ends $\lfloor t\,h_f\rfloor$. Running the same two passes with this pair as the state proves all results for every real $t\ge1$ (Section~\ref{sec:realt}).

\paragraph{Step 5 and the final bound.} Given $\bar y$, we output a cheapest assignment of the points to $F$ whose load matrix is $\bar y$. This is a transportation problem, solvable exactly in polynomial time. Fairness depends only on the loads, so the output is exactly fair whichever tree was sampled. For the cost, $\sigma$ and $\gamma$ are fixed before $T$ is sampled, so $\E[W_T(y,y^\gamma)]\le\sum_p\E[\tau(\sigma(p),\gamma(p))]\le O(\log k)\sum_p\rho(\sigma(p),\gamma(p))$, which is $O(\log k)\cdot2^{O(z)}(1+\alpha_z)\,\OPT$. Step~4 loses a further factor $4$, and returning from $\rho$ to $d^{z}$ a factor $k^{z-1}$. This gives $O(\log k)$ for $k$-median and $2^{O(z)}k^{z-1}\log k$ in general. 

\subsection{Related work}
\label{sec:related}

\paragraph{Balance and fairlets.}
\citet{ChierichettiKLV17} introduced balanced clustering for two groups and the
fairlet technique: decompose the points into small balanced groups, contract each to a
representative, and cluster the representatives with any unconstrained algorithm. They
obtained a constant-factor approximation for balanced $k$-center and an $O(t)$-approximation for balanced $k$-median.
\citet{BackursIOSVW19} sped up fairlet computation to near-linear time in the Euclidean
setting for $k$-median, and \citet*{SchmidtSS19} developed fair coresets and streaming algorithms.
\citet{BohmFLS20} handled the fully balanced multi-group case, where every cluster must
contain the same number of points from each group, obtaining constant factors for $k$-median, $k$-means and $k$-center through a reduction to the unconstrained problem; this is the case $t=1$ of Definition~\ref{def:pfc}.

\paragraph{Representational fairness.}
\citet{RosnerS18} required each cluster to reproduce the global group proportions
exactly and gave a constant factor for $k$-center. \citet{BeraCFN19} and
\citet{BerceaGKKRSS19}, independently, introduced the interval form with per-group
bounds $\alpha_i,\beta_i$ and gave constant-factor approximations for $k$-median and
$k$-means at the price of a small additive violation of the constraints.
\citet{AhmadianEKM19} treated the upper-bound-only version of $k$-center. \citet*{DaiMV22} obtained an $O(\log k)$ approximation for $k$-median with several
protected classes in time $n^{O(\ell)}$; our Theorem~\ref{thm:intro-median} matches that
ratio for the pairwise variant while running in time polynomial in $\ell$.
The two results are not comparable as statements. Representational fairness is specified by $2\ell$ thresholds, pairwise fairness by one, and for $\ell>2$ neither constraint is a special case of the other. The comparison is one of technique: their dynamic program stores a vector of $\ell$ group counts per subtree, which is what costs $n^{O(\ell)}$, and the same design applied to pairwise fairness would cost the same. Our observation is that for pairwise fairness one integer per subtree suffices, at the price of a factor $4$. Whether representational fairness admits a similarly small summary is open.

\paragraph{Pairwise fairness.}
The notion in Definition~\ref{def:pfc} appears in \citet*{BandyapadhyayFS24} under the
name $(t,k)$-fair clustering, where it is solved to a constant factor in time
$(k\ell)^{O(k)}n^{O(1)}$ using coresets. \citet{BandyapadhyayCFJ25} gave the first
polynomial-time constant factor for the $k$-center objective with arbitrary $\ell$ and
integer $t$, along with the $\Omega(k)$ integrality gap for fractional $t$.
\citet{BandyapadhyayCFJMV26} gave the first polynomial-time exactly fair
$O(k^{2}\ell t)$ approximation for $k$-median, together with the reduction from soft
uniform capacitated $k$-median that we invoke for our tightness discussion and a
strong $n^{1-\eps}$ inapproximability result for the variant with overlapping groups.

\paragraph{Other fairness notions in clustering.}
The area is large and active; \citet*{ChhabraMM21} survey it. Center-representation fairness asks
that the selected centers be diverse \citep{KleindessnerAM19,JiaSS20,AneggAKZ20};
proportional fairness asks that no large enough coalition be able to deviate profitably
\citep{ChenFLM19,MichaS20}; socially fair or minimax clustering minimizes the worst
group's average cost rather than the total
\citep{AbbasiBV21,GhadiriSV21,MakarychevV21,ChlamtacMV22,anand2026socially}; individual fairness asks that each point have a
center within a radius determined by its own neighborhood \citep{MahabadiV20,ChakrabartyN21,VakilianY22,bateni2024scalable,ebbens2024subquadratic}; fair range clustering \citep{HotegniMV23,NguyenNJ22}
and probabilistic fair clustering \citep{EsmaeiliBTD20,EsmaeiliBSD21,EsmaeiliDDB22,DickersonEMZ23,DickersonEMZ25} relax the
input assumptions.

\paragraph{Capacitated clustering.}
Our problem sits close to capacitated $k$-median, where each center may serve at most
$u$ clients. No constant-factor polynomial-time approximation is known without
violating either the capacities or the number of centers
\citep{Li17,ByrkaFRS15}; the best guarantee that violates nothing is a folklore
$O(\log k)$ obtained by probabilistic tree embedding followed by a dynamic program on
the tree \citep{CharikarCGG98,Bartal96}, and closing the gap to a constant is a
long-standing open problem \citep{CohenAddadL19,CohenAddad20}. Constant factors are
known in fixed-parameter time \citep{AdamczykBMMW19,CohenAddadL19}. Lower bounds on
cluster sizes have been studied under the name of anonymity \citep{AggarwalPFTKKZ10}. The reduction of \citet{BandyapadhyayCFJMV26} makes the connection to our problem formal in one direction.

\paragraph{Unconstrained clustering.}
We use an unconstrained approximation as a black box, so it is worth recording what is
available. For $k$-center the greedy algorithm is a $2$-approximation and this is best
possible \citep{Gonzalez85,HochbaumS85}. For $k$-median in general metrics constant
factors go back to \citet{CharikarGTS02}, with $2.675+\eps$ due to \citet{ByrkaPRST15}; local search gives $3+2/p$ with $p$ swaps \citep{AryaGKMMP04}, and the problem is hard to approximate below $1+2/e$ \citep{GuhaK99}. The line did not stop there: \citet{LiS13} gave $1+\sqrt{3}+\eps$ by
pseudo-approximation, \citet{CohenAddadGHOS22} refined the local-search analysis,
\citet{GowdaPST23} improved bi-point rounding to $2.613$, and
\citet{CohenAddadGLSS25} reached $2+\eps$, which is the current state of the art in general metrics.
For $k$-means, local search gives $9+\eps$ \citep{KanungoMNPSW04} and this barrier stood for two decades and was broken only recently by \citet{CharikarCGGLW25}, who give $\approx5.83$ by adapting a Lagrangian-multiplier-preserving greedy algorithm to squared connection costs. Approximating $k$-means below $1+8/e \approx 3.94$ is NP-hard. For general constant $z$, local search and its variants
give constant factors depending only on $z$ \citep{GuptaT08,CohenAddadSS21}.
For our purposes the relevant summary is that $\alpha_z = 2^{O(z)}$ for every
fixed $z \ge 1$, and in particular $\alpha_1 = 2+\eps$ and $\alpha_2 < 5.83$.

\section{Preliminaries}
\label{sec:prelim}

Throughout, $(\Omega,d)$ is a metric space, $P \subseteq \Omega$ is a set of $n$ points
partitioned into disjoint groups $P_1,\dots,P_\ell$, and $N_a = |P_a|$. Centers may be
opened at any point of a set $\mathcal{F} \subseteq \Omega$ of allowed locations, which
may be all of $\Omega$, or $P$, or an explicit finite candidate set; the only thing we
require is that the unconstrained algorithm we invoke uses the same $\mathcal{F}$. We
fix an exponent $z \ge 1$ and write $\cost_z(\phi) = \sum_{p \in P} d\bigl(p,\phi(p)\bigr)^{z}$
for an assignment $\phi \colon P \to \mathcal{F}$. An assignment is \emph{fair} if every
one of its clusters $\phi^{-1}(c)$ satisfies Definition~\ref{def:pfc}. We let $\OPT$
denote the optimal value of the pairwise fair instance, attained by clusters
$C_1^\star,\dots,C_k^\star$ with centers $c_1^\star,\dots,c_k^\star$. We
assume $t \ge 1$ is an integer until Section~\ref{sec:realt}, where the assumption is removed.
For an integer $x$ and a nonempty finite set $S$ of integers we write $\dist(x,S)=\min_{s\in S}|x-s|$.

\medskip
We use the following elementary inequality repeatedly\rev{.}
\begin{observation}
[Power mean]
\label{obs:powermean}
For $a_1,\dots,a_r \ge 0$ and $z \ge 1$ we have
$\bigl(\sum_{i=1}^{r} a_i\bigr)^{z} \le r^{z-1}\sum_{i=1}^{r} a_i^{z}$. In particular
$(a+b)^{z} \le 2^{z-1}(a^{z}+b^{z})$.
\end{observation}

\subsection{Feasibility}

Pairwise fairness has a clean feasibility test, which we can therefore run before doing
any work.

\begin{observation}
\label{obs:feasible}
A pairwise fair clustering of $P$ into at most $k$ nonempty clusters exists if and only
if $\max_a N_a \le t \min_a N_a$.
\end{observation}

We assume from now on that the instance is feasible. Note that the condition forces
every group to be nonempty as soon as one of them is.

\subsection{Loads and the repair problem}

Fix a finite set $F \subseteq \mathcal{F}$ of candidate centers. A \emph{load matrix} on
$F$ is a nonnegative integer matrix $y = (y_{fa})_{f \in F, a \in [\ell]}$, where
$y_{fa}$ is the number of points of group $a$ assigned to $f$. For $R \subseteq F$ we
write $y_a(R) = \sum_{f \in R} y_{fa}$. A row $y_{f\cdot}$ is \emph{fair} if
$\max_a y_{fa} \le t \min_a y_{fa}$, and $y$ is fair if all its rows are. Any assignment
$\phi \colon P \to F$ induces a load matrix, and $\phi$ is a fair assignment exactly
when its load matrix is fair.

\subsection{Hierarchically well-separated trees}

\begin{figure}[t]
\centering
\begin{tikzpicture}[lab]
\tikzset{
  inode/.style={circle,draw=cGA!75,fill=cGA!10,line width=0.7pt,
                minimum size=12pt,inner sep=0pt},
  lnode/.style={circle,draw=cGC!80,fill=cGC!25,line width=0.7pt,
                minimum size=8pt,inner sep=0pt}}
\fill[cAcc!7,rounded corners=3pt]  (1.62,-3.40) rectangle (4.18,-2.60);
\draw[cAcc!35,rounded corners=3pt,line width=0.6pt]
                                   (1.62,-3.40) rectangle (4.18,-2.60);
\foreach \y in {0,-1.5,-3.0}
  {\draw[cRule,line width=0.5pt,dashed] (-4.35,\y) -- (4.35,\y);}
\draw[cMute!70,line width=0.7pt] (0,-0.22) -- (-2.9,-1.34);
\draw[cMute!70,line width=0.7pt] (0,-0.22) -- (0,-1.34);
\draw[cAcc,line width=1.7pt]     (0,-0.22) -- (2.9,-1.34);
\draw[cMute!70,line width=0.7pt] (-2.9,-1.66) -- (-3.6,-2.88);
\draw[cMute!70,line width=0.7pt] (-2.9,-1.66) -- (-2.2,-2.88);
\draw[cMute!70,line width=0.7pt] ( 0  ,-1.66) -- (-0.7,-2.88);
\draw[cMute!70,line width=0.7pt] ( 0  ,-1.66) -- ( 0.7,-2.88);
\draw[cMute!70,line width=0.7pt] ( 2.9,-1.66) -- ( 2.2,-2.88);
\draw[cMute!70,line width=0.7pt] ( 2.9,-1.66) -- ( 3.6,-2.88);
\node[inode] at (0,0) {};
\node[inode] at (-2.9,-1.5) {};
\node[inode] at (0,-1.5) {};
\node[inode,draw=cAcc,fill=cAcc!18] at (2.9,-1.5) {};
\foreach \x in {-3.6,-2.2,-0.7,0.7} {\node[lnode] at (\x,-3.0) {};}
\foreach \x in {2.2,3.6} {\node[lnode,draw=cAcc!85,fill=cAcc!30] at (\x,-3.0) {};}
\node[left,note] at (-4.40, 0.00) {level $0$ (root)};
\node[left,note] at (-4.40,-1.50) {level $1$};
\node[left,note] at (-4.40,-3.00) {level $2$: leaves $=F$};
\node[right,note] at (4.40,-0.75) {weight $\omega/2$};
\node[right,note] at (4.40,-2.25) {weight $\omega/4$};
\node[note,text=cAcc,anchor=south east] at (2.10,-0.46) {weight $w_u$};
\node[note,text=cAcc,anchor=west] at (3.20,-1.50) {node $u$};
\node[note,text=cAcc] at (2.90,-3.66) {$F_u$};
\end{tikzpicture}
\caption[A leveled 2-HST]{A leveled $2$-HST on $m=6$ candidate centers with two levels below the root. The leaves are the centers $F$. Every non-root node $u$ has one edge to its parent, of weight $w_u$, and the leaves below $u$ form the set $F_u$. All edges entering the same level have the same weight, and the weights halve from one level to the next.}
\label{fig:hst}
\end{figure}
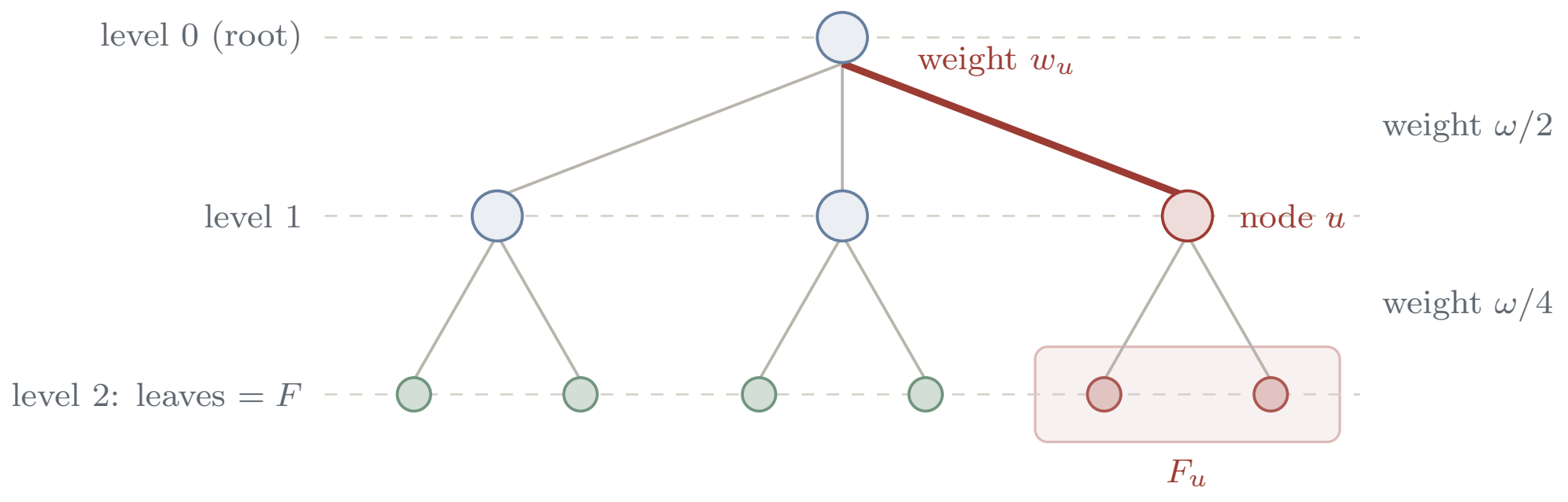

Let $T$ be a rooted tree whose leaves are the centers $F$. For a node $u$ of $T$ we write $F_u\subseteq F$ for the set of leaves in the subtree of $u$ and, when $u$ is not the root, $w_u>0$ for the weight of the edge joining $u$ to its parent. For the root we set $w_{\rt}=0$, so that a sum over all nodes of terms weighted by $w_u$ is a sum over all edges. The \emph{level} of a node is its distance from the root, counted in edges. We call $T$ a \emph{leveled $2$-HST} if all leaves are at the same level and there is a number $\omega>0$ such that $w_u=\omega\,2^{-j}$ for every node $u$ at level $j\ge1$. In particular all children of a node have the same weight, and a child of a non-root node $u$ has weight $w_u/2$. This halving is the only property of the weights that our analysis uses. We write $\tau(f,g)$ for the distance between the leaves $f,g$ in $T$, and for a load matrix $y$ on $F$ we abbreviate $y_a(u)=y_a(F_u)$, the number of points of group $a$ assigned to centers below $u$. Figure~\ref{fig:hst} fixes the picture.

The tree path between two leaves $f$ and $g$ uses the edge above $u$ exactly when one of $f,g$ lies in $F_u$ and the other does not. So $w_u$ is the price that a point pays for being reassigned from a center inside $F_u$ to a center outside it, or the other way round; we then say that the point \emph{crosses} the edge above $u$.

Finally, $m = |F|$ denotes the number of \emph{leaves} of $T$, equivalently the number
of candidate centers. In our application $F$ is the output of an unconstrained
clustering algorithm, so $m \le k$.

\begin{theorem}[\citet*{FakcharoenpholRT04}; see also \citet{Bartal96,Bartal98}]
\label{thm:frt}
There is a universal constant $C$ and a polynomial-time randomized procedure that,
given an $m$-point metric $(F,\rho)$, samples a leveled $2$-HST $T$ with leaf set $F$
such that
\begin{equation}
\label{eq:frt}
  \rho(f,g) \;\le\; \tau(f,g)
  \qquad\text{and}\qquad
  \E\bigl[\tau(f,g)\bigr] \;\le\; C\log(m+1)\,\rho(f,g)
  \qquad (f,g \in F).
\end{equation}
The number of levels is $L = O(\log \Delta)$, where $\Delta$ is the aspect ratio of
$\rho$.
\end{theorem}

\begin{remark}
\label{rem:levels}
Note that $|T| = O(mL)$, where $L = O(\log\Delta)$ is the number of levels and $\Delta$ is the aspect ratio of $\rho$, the ratio of its largest to its smallest distance between distinct centers. Since $\rho\le d^{z}$ and every $\rho$-distance is at least the smallest $d^{z}$, we have $\log\Delta\le z$ times the logarithm of the aspect ratio of $d$ on $F$, which is at most linear in the bit length of the input distances. So $|T|$ is polynomial in the input size, and this is all we need.
\end{remark}

The following standard fact explains why trees are convenient.

\begin{observation}
\label{obs:treeemd}
Let $T$ be a tree with leaf set $F$ in which every non-root node $u$ carries the weight $w_u$ of the edge to its parent, and let $x,x'$ be
nonnegative integer vectors on $F$ with $\sum_f x_f = \sum_f x'_f$. The minimum cost of
transporting $x$ to $x'$ under the tree metric $\tau$ equals
$\sum_{u} w_u \, |x(F_u) - x'(F_u)|$, the sum being over the non-root nodes, and an integral optimal solution can be computed in polynomial time.
\end{observation}
It is the observation underlying
the embedding of the earth-mover distance into $\ell_1$ via hierarchical decompositions,
due to \citet{Charikar02} and \citet{IndykT03}, and the same bottom-up routing on an HST
is what \citet{BackursIOSVW19,DaiMV22} use in their algorithms for fair clustering.
\begin{proof}[Proof of Observation~\ref{obs:treeemd}]
\emph{Lower bound.} In any transport plan, the amount moved out of $F_u$ minus the amount moved into $F_u$ equals $x(F_u)-x'(F_u)$. So at least $|x(F_u)-x'(F_u)|$ units cross the edge above $u$, each paying $w_u$.
\emph{Upper bound.} Send across the edge above every non-root node $u$ exactly $x(F_u)-x'(F_u)$ units, upward if this number is positive and downward if it is negative. This is a flow on $T$: at an internal node the amounts arriving from the children add up to the amount sent to the parent, because $F_u$ is the disjoint union of the leaf sets of the children of $u$, and nothing has to be sent above the root because the totals of $x$ and $x'$ agree. The flow carries the supply $x$ to the demand $x'$, uses the edge above $u$ exactly $|x(F_u)-x'(F_u)|$ times, and is integral. Decomposing it into leaf-to-leaf paths gives an integral transport plan of the stated cost.
\end{proof}
Accordingly, for load matrices $y,\bar y$ on $F$ with matching per-group totals we set
\begin{equation}
\label{eq:WT}
  W_T(y,\bar y) \;=\;
  \sum_{a=1}^{\ell} \sum_{u} w_u \bigl|y_a(u) - \bar y_a(u)\bigr| ,
\end{equation}
the total, over groups, of the tree transport cost of moving $y$ to $\bar y$ while
preserving group membership.

\section{Reduction to a repair problem on $k$ centers}
\label{sec:reduction}

This section shows that we may fix the center set once and for all, and then think only
about moving points among those centers. The key point is that
pairwise fairness survives merging, which is what makes the reduction possible at all.

Run any polynomial-time $\alpha_z$-approximation for unconstrained $(k,z)$-clustering on
$P$, let $F$ be its set of distinct centers, put $m = |F| \le k$, and let $\sigma(p)$ be
a nearest center of $F$ to $p$. Define
\begin{equation}
\label{eq:U}
  U \;=\; \cost_z(\sigma) \;=\; \sum_{p \in P} d\bigl(p,\sigma(p)\bigr)^{z}
  \;\le\; \alpha_z \cdot \OPT ,
\end{equation}
the inequality because every fair solution is in particular an unconstrained solution,
so the unconstrained optimum is at most $\OPT$.

\begin{observation}[Fairness is closed under union]
\label{obs:union}
If $C$ and $C'$ are pairwise fair then so is $C \cup C'$, provided $C \cap C' =
\emptyset$.
\end{observation}

\begin{proof}
$|(C\cup C')\cap P_i| = |C\cap P_i| + |C'\cap P_i|
 \le t|C\cap P_j| + t|C'\cap P_j| = t|(C\cup C')\cap P_j|$.
\end{proof}

\begin{lemma}
\label{lem:projection}
There is a fair assignment $\gamma \colon P \to F$ with
\begin{equation}
\label{eq:B}
  B \;:=\; \cost_z(\gamma)
  \;\le\; 2^{2z-1}\,\OPT + 2^{z-1}\,U
  \;\le\; \bigl(2^{2z-1} + 2^{z-1}\alpha_z\bigr)\OPT .
\end{equation}
For $z=1$ this reads $B \le 2\,\OPT + U$.
\end{lemma}

\begin{proof}
For each optimal fair cluster $C_r^\star$ let $f_r \in F$ be a center nearest to
$c_r^\star$, and let $\gamma$ send every point of $C_r^\star$ to $f_r$. Each cluster of
$\gamma$ is a disjoint union of optimal fair clusters, hence fair by
Observation~\ref{obs:union} applied inductively.

For the cost, take $p \in C_r^\star$. Since $f_r$ is a nearest point of $F$ to
$c_r^\star$ and $\sigma(p) \in F$,
\[
  d(p,f_r) \le d(p,c_r^\star) + d(c_r^\star,f_r)
           \le d(p,c_r^\star) + d(c_r^\star,\sigma(p))
           \le 2d(p,c_r^\star) + d(p,\sigma(p)).
\]
Raising to the power $z$ and applying Observation~\ref{obs:powermean} with the two
terms $2d(p,c_r^\star)$ and $d(p,\sigma(p))$ gives
$d(p,f_r)^{z} \le 2^{2z-1}d(p,c_r^\star)^{z} + 2^{z-1}d(p,\sigma(p))^{z}$. Summing over
$p$ and using \eqref{eq:U} gives \eqref{eq:B}.
\end{proof}

What Lemma~\ref{lem:projection} buys us is the guarantee that a cheap fair assignment
\emph{into $F$} exists, so in the rest of the paper, the goal is to find one.

\paragraph{Making the cost a metric.} For $z=1$ the function $d$ restricted to $F$ is already a metric and we may set
$\rho = d$. For $z>1$ the cost $d(f,g)^{z}$ violates the triangle inequality, and we
replace it by its shortest-path closure
\begin{equation}
\label{eq:rho}
  \rho(f,g) \;=\;
  \min_{f = v_0, v_1, \dots, v_r = g}\ \sum_{i=0}^{r-1} d(v_i,v_{i+1})^{z},
\end{equation}
the minimum over paths with all $v_i \in F$. This is a metric on $F$, computable in
polynomial time by all-pairs shortest paths.

\begin{lemma}[Metricization]
\label{lem:metric}
For all $f,g \in F$, $\rho(f,g) \le d(f,g)^{z} \le (m-1)^{z-1}\,\rho(f,g)$.
For $z=1$ both hold with equality and no loss. Here, $m = |F| \le k$.
\end{lemma}

\begin{proof}
The first inequality follows by taking the single-edge path. For the second, take an
optimal path realizing $\rho(f,g)$; it may be chosen simple, so it has $r \le m-1$
edges. The triangle inequality for $d$ gives
$d(f,g) \le \sum_{i} d(v_i,v_{i+1})$, and Observation~\ref{obs:powermean} then gives
$d(f,g)^{z} \le r^{z-1}\sum_i d(v_i,v_{i+1})^{z} = r^{z-1}\rho(f,g)
\le (m-1)^{z-1}\rho(f,g)$.
\end{proof}

\begin{remark}[The factor is the price of any metric]
\label{rem:metrictight}
No other choice of metric does better. Let $F=\{0,1,\dots,m-1\}\subset\R$ with $d(i,j)=|i-j|$, and let $\rho'$ be any metric on $F$ with $\lambda\rho'\le d^{z}\le\mu\rho'$ pointwise, for some $\lambda,\mu>0$. Then
\[
  (m-1)^{z}=d(0,m-1)^{z}\le\mu\,\rho'(0,m-1)\le\mu\sum_{i=0}^{m-2}\rho'(i,i+1)\le\frac{\mu}{\lambda}\sum_{i=0}^{m-2}d(i,i+1)^{z}=\frac{\mu}{\lambda}\,(m-1),
\]
so $\mu/\lambda\ge(m-1)^{z-1}$.
\end{remark}

\section{The scalar base of pairwise fairness}
\label{sec:scalar}
This section isolates the structural fact that the algorithm rests on. Throughout it, $t\ge1$ is an integer; Section~\ref{sec:realt} removes this assumption.

\begin{definition}[Windows, bases, compatibility]
\label{def:base}
For an integer $b\ge0$, the \emph{window} of $b$ is the set $I(b)=\{b,b+1,\dots,tb\}$; in particular $I(0)=\{0\}$. A \emph{base vector} $h=(h_f)_{f\in F}$ assigns to every center $f$ a nonnegative integer $h_f$, the \emph{base} of $f$. A load matrix $y'$ on $F$ is \emph{compatible} with $h$ if $y'_{fa}\in I(h_f)$ for every center $f$ and every group $a$.
\end{definition}

A base is best read as a quota: it commits the center $f$ to receiving at least $h_f$ points from every group and at most $t\,h_f$ from any, and within this window the individual counts are free. A center with base $0$ receives no one, so choosing a base vector also determines which centers are used. For example, with $\ell=3$ groups and $t=2$, a center serving $(3,5,4)$ points of the three groups is compatible with base $3$, whose window is $\{3,4,5,6\}$, and it is fair. A center serving $(2,5,4)$ is not fair: its smallest count is $2$, and $5$ lies outside the window $\{2,3,4\}$.

\begin{lemma}
\label{lem:base}
Let $t \ge 1$ be an integer.
\begin{enumerate}
\item[(i)] A vector $v \in \Z_{\ge 0}^{\ell}$ satisfies $\max_a v_a \le t\min_a v_a$ if
and only if there is an integer $h \ge 0$ with $v_a \in I(h)$ for all $a$; one may take
$h = \min_a v_a$.
\item[(ii)] (Windows add.) Let $b_1,\dots,b_s$ be nonnegative integers with sum $b$. If $x_i\in I(b_i)$ for every $i$, then $x_1+\dots+x_s\in I(b)$. Conversely, every $x\in I(b)$ can be written as $x=x_1+\dots+x_s$ with $x_i\in I(b_i)$ for every $i$. In other words, $I(b_1)+\dots+I(b_s)=I(b_1+\dots+b_s)$, where a sum of sets is the set of all sums of one element from each (the Minkowski sum).
\end{enumerate}
\end{lemma}

\begin{proof}
(i) If $\max_a v_a \le t\min_a v_a$, set $h = \min_a v_a$; then $h \le v_a$ by
definition and $v_a \le \max_a v_a \le th$. Conversely if $v_a \in I(h)$ for all $a$
then $\max_a v_a \le th \le t \min_a v_a$.

(ii) If $b_i\le x_i\le t\,b_i$ for every $i$, then adding up gives $b\le\sum_ix_i\le tb$. For the converse let $x\in I(b)$ and start with $x_i=b_i$ for all $i$; the sum is $b\le x$. As long as the sum is smaller than $x$, pick one index $i$ with $x_i<t\,b_i$ and raise this single $x_i$ by one. Every step raises the sum by exactly one, so the sum cannot jump over $x$. And a step is always available while the sum is smaller than $x$: if every $x_i$ had reached $t\,b_i$, the sum would be $tb\ge x$. So the procedure ends with the sum equal to $x$, and every $x_i$ is still in $I(b_i)$.
\end{proof}

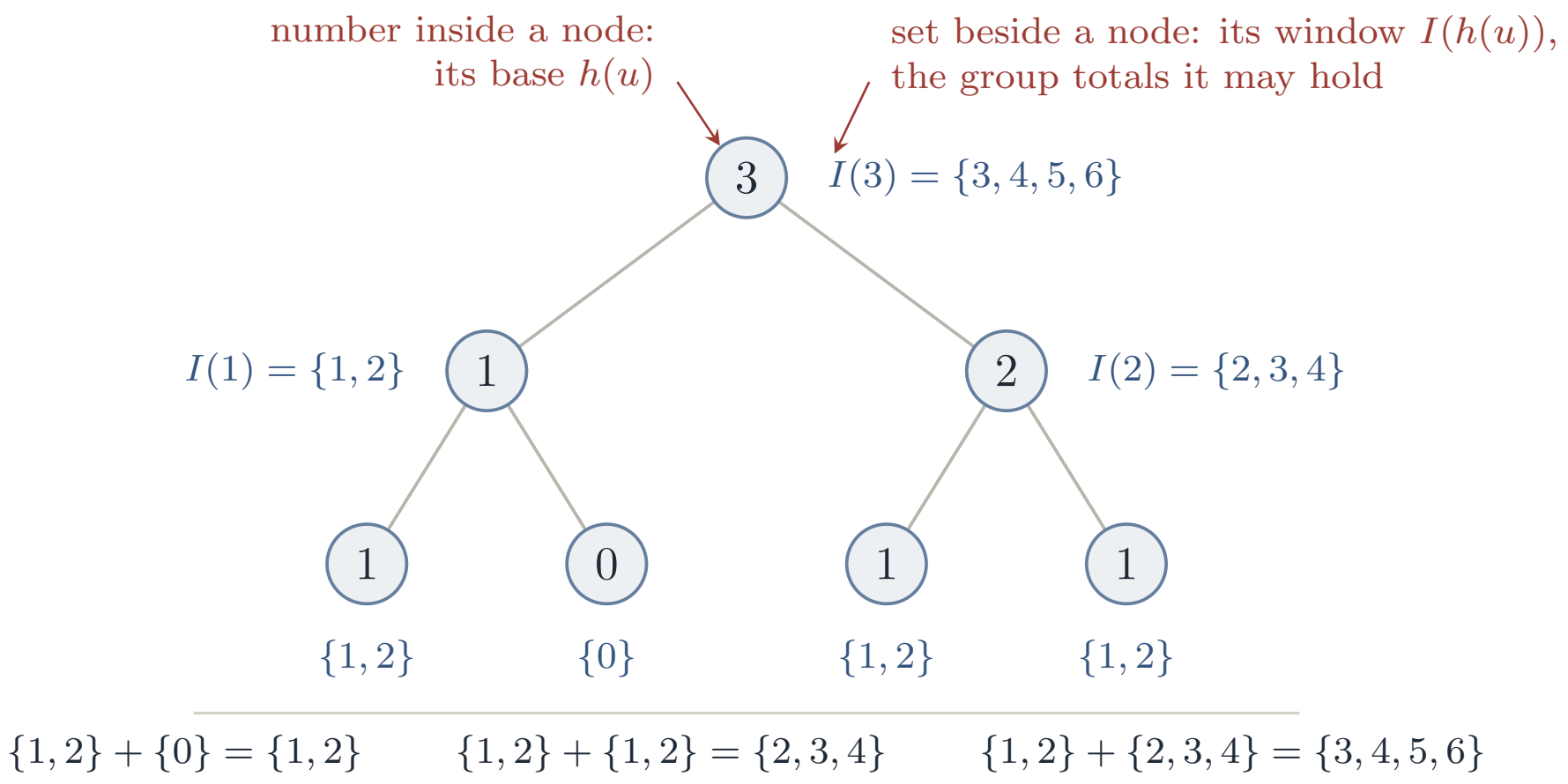
\begin{figure}[t]
\centering
\begin{tikzpicture}[lab]
\foreach \a/\b in {{0,0}/{-1.95,-1.45},{0,0}/{1.95,-1.45},
                   {-1.95,-1.45}/{-2.85,-2.90},{-1.95,-1.45}/{-1.05,-2.90},
                   {1.95,-1.45}/{1.05,-2.90},{1.95,-1.45}/{2.85,-2.90}}
  {\draw[cMute!70,line width=0.7pt] (\a) -- (\b);}
\node[tnode] (rt) at (0,0) {$3$};
\node[tnode] (R1) at (-1.95,-1.45) {$1$};
\node[tnode] (R2) at ( 1.95,-1.45) {$2$};
\node[tnode] (f1) at (-2.85,-2.90) {$1$};
\node[tnode] (f2) at (-1.05,-2.90) {$0$};
\node[tnode] (f3) at ( 1.05,-2.90) {$1$};
\node[tnode] (f4) at ( 2.85,-2.90) {$1$};
\node[note,text=cGA,right=1.6mm of rt] {$I(3)=\{3,4,5,6\}$};
\node[note,text=cGA,left=1.6mm of R1]  {$I(1)=\{1,2\}$};
\node[note,text=cGA,right=1.6mm of R2] {$I(2)=\{2,3,4\}$};
\node[note,text=cGA,below=1.2mm of f1] {$\{1,2\}$};
\node[note,text=cGA,below=1.2mm of f2] {$\{0\}$};
\node[note,text=cGA,below=1.2mm of f3] {$\{1,2\}$};
\node[note,text=cGA,below=1.2mm of f4] {$\{1,2\}$};
\node[note,text=cAcc,anchor=east,align=right] at (-0.55,0.92)
  {number inside a node:\\ its base $h(u)$};
\draw[cAcc,line width=0.5pt,->] (-0.52,0.72) -- (-0.20,0.24);
\node[note,text=cAcc,anchor=west,align=left] at (0.95,0.92)
  {set beside a node: its window $I(h(u))$,\\ the group totals it may hold};
\draw[cAcc,line width=0.5pt,->] (0.92,0.72) -- (0.66,0.18);
\draw[cRule,line width=0.6pt] (-4.15,-4.02) -- (4.15,-4.02);
\node[note,text=cInk] at (0,-4.32)
  {$\{1,2\}+\{0\}=\{1,2\}$ \qquad $\{1,2\}+\{1,2\}=\{2,3,4\}$
   \qquad $\{1,2\}+\{2,3,4\}=\{3,4,5,6\}$};
\end{tikzpicture}
\caption[Bases on a tree]{Bases on a tree of four centers, for any number $\ell$ of groups, with $t=2$.
\emph{The number written inside a node is a base}: at a leaf, the base $h_f$ chosen for that center; at an internal node, the sum $h(u)$ of the bases of the leaves below it ($1+0=1$ on the left, $1+1=2$ on the right, $1+2=3$ at the root).
\emph{The set written beside a node is its window} $I(h(u))$, the values that the number of points of any one group inside that subtree may take. The windows of the children add up to the window of their parent, as the bottom line spells out.}
\label{fig:base}
\end{figure}

\paragraph{Bases on a tree.} Let the centers be the leaves of a tree $T$, let $h$ be a base vector, and define the base of a node as the total base of the leaves below it, $h(u)=\sum_{f\in F_u}h_f$. We use Lemma~\ref{lem:base}(ii) in two directions, illustrated in Figure~\ref{fig:base}.
\begin{itemize}[leftmargin=*]
\item \emph{Upward.} If a load matrix $y'$ is compatible with $h$, then $y'_a(u)\in I(h(u))$ for every node $u$ and every group $a$, because $y'_a(u)$ is a sum of one number from the window of each leaf below $u$.
\item \emph{Downward.} If a node $u$ with children $u_1,\dots,u_s$ is handed a number $x\in I(h(u))$, it can split it as $x=x_1+\dots+x_s$ with $x_i\in I(h(u_i))$, because $h(u)=\sum_ih(u_i)$.
\end{itemize}
So the set of group totals that a subtree can hold, when all its leaves are to be compatible with $h$, is exactly $I(h(u))$: one window, determined by one integer, whatever the number of groups and however the base is distributed among the leaves. Group totals of subtrees are also all that the transport cost \eqref{eq:WT} depends on. This is why one integer per node can serve as the state of a dynamic program. Note that the base of an internal node is a sum of leaf bases and not the smallest group total below it: leaves with loads $(1,2)$ and $(2,1)$ have row minimum $1$ each, so with these bases their parent has base $2$, although its group totals are $(3,3)$.

\begin{remark}
Integrality of $t$ is used only in Lemma~\ref{lem:base}(ii). For fractional $t$ the
window $\{b,\dots,\lfloor tb\rfloor\}$ still absorbs sums in the direction we need for
the comparison step, but the reverse inclusion can fail, so a base alone no longer
certifies what a subtree can realize. Section~\ref{sec:realt} repairs this with a
two-parameter state.
\end{remark}

\section{Repairing loads on a tree}
\label{sec:repair}

We come to the main technical part. Throughout this section $t\ge1$ is an integer, $T$ is a leveled $2$-HST with leaf set $F$, and $y$ is a nonnegative integral load matrix on $F$ whose group totals $N_a=\sum_fy_{fa}$ satisfy $\max_aN_a\le t\min_aN_a$; we write $n=\sum_aN_a$. The \emph{repair problem} asks for a fair load matrix $\bar y$ with the same group totals as $y$ that minimizes the tree transport cost $W_T(y,\bar y)$ of \eqref{eq:WT}. We do not solve it exactly. We find a solution within a factor $4$ of the optimal solution, using one integer of state per tree node.

\paragraph{The surrogate.} Fix a base vector $h$ and recall that $h(u)=\sum_{f\in F_u}h_f$. For a node $u$, a group $a$ and an integer $b\ge0$ let
\[
  \delta^a_u(b)\;=\;\dist\bigl(y_a(u),I(b)\bigr)\;=\;\max\bigl\{0,\ b-y_a(u),\ y_a(u)-tb\bigr\},
  \qquad
  \delta_u(b)\;=\;\sum_{a=1}^{\ell}\delta^a_u(b).
\]
If the subtree of $u$ is to end up with base $b$, its group-$a$ total must move from $y_a(u)$ into the window $I(b)$, so at least $\delta^a_u(b)$ points of group $a$ must cross the edge above $u$. We call $\delta_u(b)$ the \emph{forced traffic} at $u$ under base $b$, and we define the \emph{surrogate cost} of $h$ as
\begin{equation}
\label{eq:Phi}
  \Phi_T(h)\;=\;\sum_{u}w_u\,\delta_u\bigl(h(u)\bigr),
\end{equation}
the sum being over all nodes of $T$ (recall that $w_{\rt}=0$). We call $h$ \emph{admissible} if the window of the root contains every group total, $N_a\in I(h(\rt))$ for all $a$, that is, if
\begin{equation}
\label{eq:rootcond}
  \Bigl\lceil\frac{\max_aN_a}{t}\Bigr\rceil\;\le\;\sum_{f\in F}h_f\;\le\;\min_aN_a .
\end{equation}
The range in \eqref{eq:rootcond} is nonempty exactly because $\max_aN_a\le t\min_aN_a$.

\begin{lemma}[The surrogate is a lower bound]
\label{lem:lower}
Let $h$ be a base vector, and let $y'$ be a load matrix that is compatible with $h$ and has the same group totals as $y$. Then $h$ is admissible and $\Phi_T(h)\le W_T(y,y')$.
\end{lemma}

\begin{proof}
By Lemma~\ref{lem:base}(ii), applied to the leaves below a node $u$, compatibility gives $y'_a(u)\in I(h(u))$ for every node $u$ and every group $a$. At the root this reads $N_a\in I(h(\rt))$, so $h$ is admissible. At every node it gives $\delta^a_u(h(u))=\dist\bigl(y_a(u),I(h(u))\bigr)\le|y_a(u)-y'_a(u)|$. Multiplying by $w_u$ and summing over all nodes and groups gives $\Phi_T(h)\le W_T(y,y')$.
\end{proof}

\paragraph{Pass 1 (bottom-up): choosing the bases.} For a node $u$ and an integer $b\in\{0,1,\dots,n\}$ let $V_u(b)$ be the smallest value of $\sum_{u'}w_{u'}\,\delta_{u'}(h(u'))$, where $u'$ ranges over the nodes of the subtree of $u$ (including $u$), over all choices of bases for the leaves below $u$ with $h(u)=b$. For a leaf $f$ there is one choice, $h_f=b$, and $V_f(b)=w_f\,\delta_f(b)$. For an internal node $u$ with children $u_1,\dots,u_s$,
\begin{equation}
\label{eq:dpup}
  V_u(b)\;=\;w_u\,\delta_u(b)\;+\;
  \min_{\substack{b_1+\cdots+b_s\,=\,b\\ b_i\in\Z_{\ge0}}}\ \sum_{i=1}^{s}V_{u_i}(b_i),
\end{equation}
because the term of $u$ itself depends on the bases below it only through their sum $b$, and the bases in different children can be chosen independently once $b$ has been split. Tables indexed by $b\le n$ suffice: bases are nonnegative, so an admissible base vector has $h(u)\le h(\rt)\le\min_aN_a\le n$ at every node. At the root we take a $b$ in the range \eqref{eq:rootcond} that minimizes $V_{\rt}(b)$, and we recover a base vector $h$ by backtracking through \eqref{eq:dpup}. By construction $h$ is admissible and
\begin{equation}
\label{eq:dpopt}
  \Phi_T(h)\;\le\;\Phi_T(h')\qquad\text{for every admissible base vector }h'.
\end{equation}

\paragraph{Pass 2 (top-down): from bases to loads.} Let $h$ be any admissible base vector. We treat each group $a$ separately and give every node $u$ a \emph{target} $x_a(u)\in I(h(u))$, the number of points of group $a$ that the subtree of $u$ will hold after the repair. The root gets $x_a(\rt)=N_a$, which lies in its window because $h$ is admissible. Let $u$ be an internal node that has received its target, with children $u_1,\dots,u_s$.
\begin{enumerate}[leftmargin=*]
\item \emph{Offer.} Give each child the element of its own window closest to what it currently holds,
\begin{equation}
\label{eq:dpdown}
  q_i\;=\;\min\bigl\{\max\{y_a(u_i),\,h(u_i)\},\ t\,h(u_i)\bigr\}\qquad(1\le i\le s),
\end{equation}
so that $|q_i-y_a(u_i)|=\delta^a_{u_i}(h(u_i))$.
\item \emph{Correct.} Let $D=x_a(u)-\sum_iq_i$. If $D=0$, the offers are the targets, $x_a(u_i)=q_i$. If $D>0$, the offers fall short of the target by $D$, and we hand out the missing amount to the children \emph{one after another}, in the order $i=1,\dots,s$, each child taking as much as its window allows. Writing $D_i$ for the amount still missing when child $i$ is reached, with $D_1=D$,
\[
  \theta_i=\min\bigl\{D_i,\ t\,h(u_i)-q_i\bigr\},\qquad
  x_a(u_i)=q_i+\theta_i,\qquad
  D_{i+1}=D_i-\theta_i .
\]
If $D<0$, the offers exceed the target by $|D|$, and we take the excess back in the same way: $D_1=|D|$, $\theta_i=\min\{D_i,\ q_i-h(u_i)\}$, $x_a(u_i)=q_i-\theta_i$ and $D_{i+1}=D_i-\theta_i$.
\end{enumerate}
So the offers are changed one child at a time, and all changes at a node go in the same direction. Nothing is left over at the end. Suppose $D>0$ and $D_{s+1}>0$. Then every child was raised to the top of its window, so the targets sum to $t\,h(u)\ge x_a(u)$; but they also sum to $x_a(u)-D_{s+1}<x_a(u)$, a contradiction. The case $D<0$ is symmetric, with the bottoms of the windows and $h(u)\le x_a(u)$. The order of the children is fixed only to make the procedure deterministic. The analysis below uses just three properties, and they hold for any order: every target lies in its window, the targets sum to $x_a(u)$, and the corrections $x_a(u_i)-q_i$ are all $\ge0$ or all $\le0$. The output is $\bar y_{fa}=x_a(f)$ for every leaf $f$.

\begin{lemma}[Reconstruction]
\label{lem:recon}
For every admissible base vector $h$, Pass~2 outputs an integral load matrix $\bar y$ that has the same group totals as $y$, is compatible with $h$, that is, $h_f\le\bar y_{fa}\le t\,h_f$ for all $f\in F$ and $a\in[\ell]$, and satisfies $W_T(y,\bar y)\le4\,\Phi_T(h)$.
\end{lemma}

\begin{proof}
The targets of the children of a node sum to the target of the node, so $x_a(u)=\sum_{f\in F_u}\bar y_{fa}=\bar y_a(u)$ for every node $u$. At the root this says that $\bar y$ has the group totals $N_a$, and at the leaves $\bar y_{fa}=x_a(f)\in I(h_f)$ is compatibility.

For the cost, fix a group $a$ and write $e(u)=|x_a(u)-y_a(u)|$ for the \emph{error} of a node and $d(u)=\delta^a_u(h(u))$ for its forced traffic. Let $u$ be an internal node with children $u_1,\dots,u_s$. Then
\begin{alignat}{2}
  \sum_i e(u_i)
   &\;\le\; \sum_i\bigl|x_a(u_i)-q_i\bigr|+\sum_i\bigl|q_i-y_a(u_i)\bigr|
     &\quad& \rhd\ \text{triangle inequality through } q_i \notag\\
   &\;=\; \Bigl|x_a(u)-\sum_i q_i\Bigr|+\sum_i d(u_i)
     && \rhd\ \text{corrections share one sign} \notag\\
   &\;\le\; e(u)+\Bigl|y_a(u)-\sum_i q_i\Bigr|+\sum_i d(u_i)
     && \rhd\ \text{triangle inequality through } y_a(u) \notag\\
   &\;\le\; e(u)+2\sum_i d(u_i)
     && \rhd\ y_a(u)=\textstyle\sum_i y_a(u_i) .
  \label{eq:reconstruct}
\end{alignat}
In the second line, all corrections $x_a(u_i)-q_i$ have the same sign (as described in the correction step of the top-down pass), so their absolute values add up to the absolute value of their sum, which is $x_a(u)-\sum_iq_i$; and $|q_i-y_a(u_i)|=d(u_i)$ by the choice of $q_i$ in \eqref{eq:dpdown}. In the last line, $|y_a(u)-\sum_iq_i|\le\sum_i|y_a(u_i)-q_i|=\sum_id(u_i)$.

Now we bring in the weights. All children of $u$ have the same weight, and it equals $w_u/2$ unless $u$ is the root, in which case $e(u)=0$ because $x_a(\rt)=N_a=y_a(\rt)$. Multiplying \eqref{eq:reconstruct} by this common weight therefore gives, for every internal node $u$,
\[
  \sum_i w_{u_i}\,e(u_i)\;\le\;\tfrac12\,w_u\,e(u)+2\sum_i w_{u_i}\,d(u_i).
\]
Sum this over all internal nodes $u$, and let $E=\sum_uw_u\,e(u)$ and $D=\sum_uw_u\,d(u)$, both sums over all nodes. Every node other than the root is a child of exactly one internal node, so the left-hand sides add up to $E$, and the right-hand sides to at most $\tfrac12E+2D$. Hence $E\le\tfrac12E+2D$, that is, $E\le4D$. Finally sum over the groups: $E$ adds up to $W_T(y,\bar y)$ by \eqref{eq:WT}, since $x_a(u)=\bar y_a(u)$, and $D$ adds up to $\Phi_T(h)$ by \eqref{eq:Phi}.
\end{proof}

\begin{lemma}[Repair]
\label{lem:repair}
Let $t\ge1$ be an integer, let $T$ be a leveled $2$-HST with leaf set $F$, and let $y$ be a nonnegative integral load matrix on $F$ whose group totals $N_a$ satisfy $\max_aN_a\le t\min_aN_a$. Using only $y$ and $T$, Passes~1 and~2 compute, in time $O\bigl(|T|\,(n^{2}+n\ell)\bigr)$ where $n=\sum_aN_a$, integers $h_f\ge0$ and an integral load matrix $\bar y$ with
\[
  \sum_{f\in F}\bar y_{fa}=N_a
  \qquad\text{and}\qquad
  h_f\le\bar y_{fa}\le t\,h_f
  \qquad(f\in F,\ a\in[\ell]),
\]
so that every row of $\bar y$ is exactly fair, and such that for \emph{every} integral load matrix $y^\star$ with fair rows and the same group totals $N_a$,
\begin{equation}
\label{eq:repair}
  W_T(y,\bar y)\;\le\;4\,W_T(y,y^\star).
\end{equation}
\end{lemma}

\begin{proof}
Let $h$ be the base vector returned by Pass~1 and $\bar y$ the output of Pass~2 for this $h$. The properties of $\bar y$ other than \eqref{eq:repair} are in Lemma~\ref{lem:recon}; a row with $h_f\le\bar y_{fa}\le t\,h_f$ for all $a$ is fair by Lemma~\ref{lem:base}(i). For \eqref{eq:repair}, let $y^\star$ be as in the statement and put $h^\star_f=\min_ay^\star_{fa}$. By Lemma~\ref{lem:base}(i), $y^\star$ is compatible with $h^\star$, so by Lemma~\ref{lem:lower} the vector $h^\star$ is admissible and $\Phi_T(h^\star)\le W_T(y,y^\star)$. This is where it matters that $y^\star$ has the same group totals as $y$. Together with Lemma~\ref{lem:recon} and \eqref{eq:dpopt},
\[
  W_T(y,\bar y)\;\le\;4\,\Phi_T(h)\;\le\;4\,\Phi_T(h^\star)\;\le\;4\,W_T(y,y^\star).
\]

\emph{Running time.} Each table has $n+1$ entries. The totals $y_a(u)$ are computed bottom-up, and from them all values $\delta_u(b)$, $0\le b\le n$, in time $O(n\ell)$ per node. The minimum in \eqref{eq:dpup} is evaluated by adding the children one at a time. Combining two tables means trying, for every $b$, every way of splitting $b$ between them, which takes $O(n^{2})$ time, and the number of such combinations over the whole tree is less than the number of nodes. Pass~1 therefore takes $O(|T|(n^{2}+n\ell))$ time, with $\ell$ entering additively. Pass~2 takes time linear in $\ell\,|T|$, since the corrections at a node can be made greedily, one child after another. By Remark~\ref{rem:levels}, $|T|$ is polynomial in the input size.
\end{proof}

\begin{figure}[t]
\centering
\begin{tikzpicture}[lab]
\foreach \a/\b in {{0,0}/{-2.15,-1.45},{0,0}/{2.15,-1.45},
                   {-2.15,-1.45}/{-3.15,-2.90},{-2.15,-1.45}/{-1.15,-2.90},
                   {2.15,-1.45}/{1.15,-2.90},{2.15,-1.45}/{3.15,-2.90}}
  {\draw[cMute!70,line width=0.7pt] (\a) -- (\b);}
\node[vnode] (rt) at (0,0) {$(6,4)$};
\node[vnode] (R1) at (-2.15,-1.45) {$(2,1)$};
\node[vnode] (R2) at ( 2.15,-1.45) {$(4,3)$};
\node[vnode] (f1) at (-3.15,-2.90) {$(2,1)$};
\node[vnode] (f2) at (-1.15,-2.90) {$(0,0)$};
\node[vnode] (f3) at ( 1.15,-2.90) {$(2,2)$};
\node[vnode] (f4) at ( 3.15,-2.90) {$(2,1)$};
\node[note,text=cAcc,below=1.3mm of f1] {$(3,0)$};
\node[note,text=cAcc,below=1.3mm of f2] {$(0,1)$};
\node[note,below=1.3mm of f3] {$(1,2)$};
\node[note,below=1.3mm of f4] {$(2,1)$};
\draw[->,cGA,line width=0.9pt] (-4.40,-3.10) -- (-4.40,0.30);
\node[rotate=90,note,text=cGA,align=center] at (-4.72,-1.40)
  {Pass 1\ \ DP over $h(u)$};
\draw[->,cGB,line width=0.9pt] (4.40,0.30) -- (4.40,-3.10);
\node[rotate=-90,note,text=cGB!80!black,align=center] at (4.72,-1.40)
  {Pass 2\ \ split totals};
\end{tikzpicture}
\caption[The two passes]{The two passes of Lemma~\ref{lem:repair} on the tree of
Figure~\ref{fig:base}, with $\ell=2$ groups, $t=2$, and group totals $N=(6,4)$. The badges
below the leaves are the input loads $y$, with the two unfair leaves flagged in
orange: each has a group with no points. Pass~1 runs bottom up and picks the bases of
Figure~\ref{fig:base}. Pass~2 runs top down, handing each node a target vector whose
every coordinate lies in that node's window $I(h(u))$, and splitting it among the
children. Every leaf of the output is exactly fair and only two points move.
The bases shown minimize $\Phi_T$ on this instance, tied with two other base vectors.}
\label{fig:repair}
\end{figure}
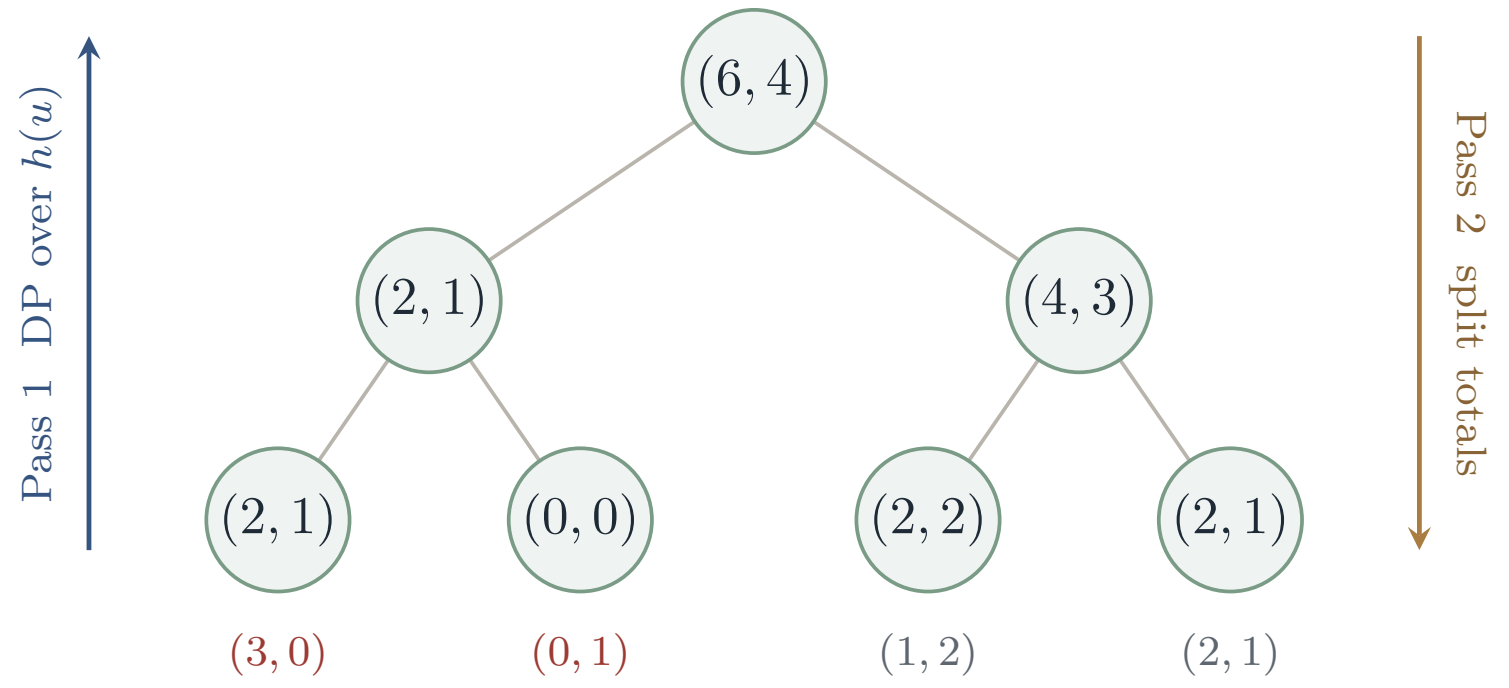

Note what the dynamic program does and does not do. It does not minimize the tree transport cost itself; that would require recording, for each group, how many points cross each edge, which is an $\ell$-dimensional state. It minimizes the surrogate $\Phi_T$. By Lemma~\ref{lem:lower} the surrogate never exceeds the cost of a fair target, and by Lemma~\ref{lem:recon} an actual repair costs at most four times the surrogate. Trading the exact objective for a one-dimensional one, at the price of this constant, is what turns $n^{O(\ell)}$ into a polynomial.

\section{The algorithm and the main theorems}
\label{sec:main}

\paragraph{Algorithm.}
The five items below are the five steps of Section~\ref{sec:overview} and the five boxes of Figure~\ref{fig:pipeline}, in the same order.
Given a feasible instance: \begin{algbox}
\begin{enumerate}
\item Run a polynomial-time $\alpha_z$-approximation for unconstrained $(k,z)$-clustering. Let $F$ be its distinct centers, $m = |F| \le k$, and let $\sigma: P \rightarrow F$ assign each point to a nearest center of $F$. If $m = 1$, return $\sigma$: the single cluster is $P$, which is fair. Otherwise continue.

\item Build the metric $\rho$ of \eqref{eq:rho} on $F$; for $z = 1$ take $\rho = d$.

\item Sample a $2$-HST $T$ on $(F,\rho)$ using Theorem~\ref{thm:frt}.

\item Let $y_{fa} = |\{p \in P_a : \sigma(p) = f\}|$ and apply Lemma~\ref{lem:repair} to $y$ and $T$, obtaining fair loads $\bar y$.

\item Return a minimum-cost assignment of $P$ to $F$ whose load matrix is exactly $\bar y$. This is a transportation problem with integral data, so an integral optimal solution exists and is computable in polynomial time \citep{Schrijver03}.
\end{enumerate}
\end{algbox}

We emphasize that fairness
is a property of the load matrix $\bar y$ alone, so \emph{every} assignment realizing
$\bar y$ is fair, and the algorithm may as well return the cheapest of them; that is
what Step 5 does. The proof of Theorem~\ref{thm:kz}, on the other hand, bounds the cost
of one particular assignment realizing $\bar y$, namely the one obtained by routing
points along the sampled tree. Since Step 5 returns a minimum-cost assignment with load
matrix $\bar y$, its cost is at most that of the routed one, so every bound proved for
the routed assignment holds for the returned one too. We implement the better
assignment and analyze the one we can bound.

\begin{theorem}[Pairwise fair $(k,z)$-clustering]
\label{thm:kz}
Let $z \ge 1$, let $t \ge 1$ be an integer, and let $\alpha_z$ be the ratio of a polynomial-time approximation for unconstrained $(k,z)$-clustering. On a feasible instance the algorithm above runs in polynomial time, opens at most $k$ centers, outputs an assignment satisfying every pairwise fairness constraint exactly, and has
\[
  \E[\cost_z]
  \;\le\;
  \Bigl[\,2^{z-1}\alpha_z
    \;+\; 2^{2z}\,C\,k^{z-1}\log(k+1)\bigl((1+2^{z-1})\alpha_z + 2^{2z-1}\bigr)\Bigr]\cdot\OPT ,
\]
where $C$ is the constant of Theorem~\ref{thm:frt}. In particular the ratio is $2^{O(z)}k^{z-1}\log k$, since $\alpha_z=2^{O(z)}$ is available (Section~\ref{sec:related}).
\end{theorem}

\begin{proof}
\emph{Feasibility of the output.} By Lemma~\ref{lem:repair} every row of $\bar y$
satisfies $h_f \le \bar y_{fa} \le t\cdot h_f$, so
$\max_a \bar y_{fa} \le t\cdot h_f \le t \min_a \bar y_{fa}$ and each nonempty cluster is
exactly fair. The group totals are preserved, so $\bar y$ is realizable by an assignment
of $P$. The number of nonempty clusters is at most $|F| = m \le k$.

\emph{Cost.} Let $y^\gamma$ be the load matrix of the comparison assignment $\gamma: P \rightarrow F$ of Lemma~\ref{lem:projection}. Both $y$ and $y^\gamma$ arise from assignments of all of $P$, so they have the same group totals $N_a$, and every row of $y^\gamma$ is fair.
Since $T$ is sampled independently of $\sigma$ and $\gamma$, and since the pairs
$(\sigma(p),\gamma(p))$ define a group-preserving coupling from $y$ to $y^\gamma$,
Observation~\ref{obs:treeemd} gives
\begin{equation}
\label{eq:couple}
  W_T(y, y^\gamma) \;\le\; \sum_{p \in P} \tau\bigl(\sigma(p),\gamma(p)\bigr),
\end{equation}
and Lemma~\ref{lem:repair} then gives
$W_T(y,\bar y) \le 4\sum_{p} \tau(\sigma(p),\gamma(p))$.

Let $q^a_{fg}$ be an integral optimal coupling from $y_a$ to $\bar y_a$ on $T$, which
exists by Observation~\ref{obs:treeemd}, and consider the assignment that moves, for
each group $a$ and each pair $f,g$, exactly $q^a_{fg}$ points of group $a$ from $f$ to
$g$. Its load matrix is $\bar y$. By Lemma~\ref{lem:metric} and the domination in
\eqref{eq:frt},
\[
  \sum_{a,f,g} q^a_{fg}\, d(f,g)^{z}
  \;\le\; (m-1)^{z-1}\sum_{a,f,g} q^a_{fg}\,\rho(f,g)
  \;\le\; (m-1)^{z-1}\sum_{a,f,g} q^a_{fg}\,\tau(f,g)
  \;=\; (m-1)^{z-1} W_T(y,\bar y).
\]
A point $p$ moved from $f = \sigma(p)$ to $g$ satisfies
$d(p,g)^{z} \le 2^{z-1}\bigl(d(p,f)^{z} + d(f,g)^{z}\bigr)$ by
Observation~\ref{obs:powermean}. Summing,
\begin{equation}
\label{eq:preexp}
  \cost_z \;\le\; 2^{z-1}U + 2^{z-1}(m-1)^{z-1} W_T(y,\bar y)
       \;\le\; 2^{z-1}U + 2^{z+1}(m-1)^{z-1}\sum_{p\in P}\tau\bigl(\sigma(p),\gamma(p)\bigr).
\end{equation}
Now take expectations. The assignments $\sigma$ and $\gamma$ are fixed before $T$ is
sampled, so by \eqref{eq:frt} and then Lemma~\ref{lem:metric},
\[
  \E\Bigl[\sum_p \tau\bigl(\sigma(p),\gamma(p)\bigr)\Bigr]
  \;\le\; C\log(m+1)\sum_p \rho\bigl(\sigma(p),\gamma(p)\bigr)
  \;\le\; C\log(m+1)\sum_p d\bigl(\sigma(p),\gamma(p)\bigr)^{z}.
\]
Routing through $p$ and applying Observation~\ref{obs:powermean} once more,
$\sum_p d(\sigma(p),\gamma(p))^{z} \le 2^{z-1}(U + B)$. Combining with
\eqref{eq:preexp},
\[
  \E[\cost_z]
   \;\le\; 2^{z-1}U + 2^{2z}\,C\,(m-1)^{z-1}\log(m+1)\,(U+B).
\]
Finally $U \le \alpha_z\OPT$ by \eqref{eq:U}, $B \le (2^{2z-1}+2^{z-1}\alpha_z)\OPT$ by
\eqref{eq:B}, so $U + B \le \bigl((1+2^{z-1})\alpha_z + 2^{2z-1}\bigr)\OPT$, and
$m \le k$. So, the claimed bound follows.

\emph{Running time.} Steps 1, 2, 3, and 5 are standard polynomial-time computations, and Step 4 is polynomial by Lemma~\ref{lem:repair} and Remark~\ref{rem:levels}.
\end{proof}

Setting $z=1$, where Lemma~\ref{lem:metric} is lossless and
$(m-1)^{z-1} = 1$, gives the main $k$-median result.

\begin{theorem}[Pairwise fair $k$-median]
\label{thm:median}
Let $t \ge 1$ be an integer and let $\alpha$ be the ratio of a polynomial-time
approximation for unconstrained $k$-median. On a feasible instance the algorithm above
runs in polynomial time, opens at most $k$ centers, is exactly fair, and satisfies $\E[\cost_1] = O(\log k)\cdot \OPT$.
\end{theorem}

\begin{corollary}[Pairwise fair $k$-means]
\label{cor:means}
Under the same hypotheses with $z=2$,
$\E[\cost_2] = O(k\log k)\cdot\OPT$.
Replacing each returned center by the centroid of its cluster only decreases the cost.
\end{corollary}

Note that every output of the algorithm is exactly fair, whatever the sampled tree, so only the cost is random. By Markov's inequality, the cost of a single run exceeds twice the bound on its expectation with probability at most $1/2$. Hence the cheapest of $\lceil\log_2(1/\eta)\rceil$ independent runs costs at most twice that bound with probability at least $1-\eta$. No feasibility check beyond Observation~\ref{obs:feasible} is needed.

\subsubsection*{Generative AI Disclosure}
The results were obtained through iterative and interactive engagement between the author and AI models, including GPT-5.6 Sol and Opus 5. Opus 5 and Fable 5.2 were also used to draft and revise the paper. The author has verified the correctness and originality of all content, including references.

\bibliographystyle{abbrvnat}
\bibliography{ref}

@inproceedings{BandyapadhyayCFJ25,
  author    = {Sayan Bandyapadhyay and Tianzhi Chen and Zachary Friggstad and Mahya Jamshidian},
  title     = {A Constant-Factor Approximation for Pairwise Fair $k$-Center Clustering},
  booktitle = {Integer Programming and Combinatorial Optimization (IPCO)},
  series    = {Lecture Notes in Computer Science},
  volume    = {15620},
  pages     = {43--57},
  publisher = {Springer},
  year      = {2025}
}

@inproceedings{BandyapadhyayCFJMV26,
  author    = {Sayan Bandyapadhyay and Eden Chlamt{\'a}{\v c} and Zachary Friggstad and
               Mahya Jamshidian and Yury Makarychev and Ali Vakilian},
  title     = {A Polynomial-Time Approximation for Pairwise Fair $k$-Median Clustering},
  booktitle = {International Conference on Artificial Intelligence and Statistics (AISTATS)},
  year      = {2026}
}

@article{BandyapadhyayFS24,
  author  = {Sayan Bandyapadhyay and Fedor V. Fomin and Kirill Simonov},
  title   = {On coresets for fair clustering in metric and {E}uclidean spaces and their applications},
  journal = {Journal of Computer and System Sciences},
  volume  = {142},
  pages   = {103506},
  year    = {2024}
}

@inproceedings{ChierichettiKLV17,
  author    = {Flavio Chierichetti and Ravi Kumar and Silvio Lattanzi and Sergei Vassilvitskii},
  title     = {Fair Clustering Through Fairlets},
  booktitle = {Advances in Neural Information Processing Systems (NeurIPS)},
  pages     = {5029--5037},
  year      = {2017}
}

@article{BohmFLS20,
  author  = {Matteo B{\"o}hm and Adriano Fazzone and Stefano Leonardi and Chris Schwiegelshohn},
  title   = {Fair Clustering with Multiple Colors},
  journal = {arXiv preprint arXiv:2002.07892},
  year    = {2020}
}

@inproceedings{BackursIOSVW19,
  author    = {Arturs Backurs and Piotr Indyk and Krzysztof Onak and Baruch Schieber and
               Ali Vakilian and Tal Wagner},
  title     = {Scalable Fair Clustering},
  booktitle = {International Conference on Machine Learning (ICML)},
  pages     = {405--413},
  year      = {2019}
}

@inproceedings{BeraCFN19,
  author    = {Suman Kalyan Bera and Deeparnab Chakrabarty and Nicolas Flores and Maryam Negahbani},
  title     = {Fair Algorithms for Clustering},
  booktitle = {Advances in Neural Information Processing Systems (NeurIPS)},
  pages     = {4954--4965},
  year      = {2019}
}

@inproceedings{BerceaGKKRSS19,
  author    = {Ioana O. Bercea and Martin Gro{\ss} and Samir Khuller and Aounon Kumar and
               Clemens R{\"o}sner and Daniel R. Schmidt and Melanie Schmidt},
  title     = {On the Cost of Essentially Fair Clusterings},
  booktitle = {Approximation, Randomization, and Combinatorial Optimization (APPROX/RANDOM)},
  year      = {2019}
}

@inproceedings{RosnerS18,
  author    = {Clemens R{\"o}sner and Melanie Schmidt},
  title     = {Privacy Preserving Clustering with Constraints},
  booktitle = {International Colloquium on Automata, Languages, and Programming (ICALP)},
  year      = {2018}
}

@inproceedings{AhmadianEKM19,
  author    = {Sara Ahmadian and Alessandro Epasto and Ravi Kumar and Mohammad Mahdian},
  title     = {Clustering without Over-Representation},
  booktitle = {ACM SIGKDD International Conference on Knowledge Discovery and Data Mining (KDD)},
  pages     = {267--275},
  year      = {2019}
}

@inproceedings{DaiMV22,
  author    = {Zhen Dai and Yury Makarychev and Ali Vakilian},
  title     = {Fair Representation Clustering with Several Protected Classes},
  booktitle = {ACM Conference on Fairness, Accountability, and Transparency (FAccT)},
  pages     = {814--823},
  year      = {2022}
}

@inproceedings{SchmidtSS19,
  author    = {Melanie Schmidt and Chris Schwiegelshohn and Christian Sohler},
  title     = {Fair Coresets and Streaming Algorithms for Fair $k$-Means},
  booktitle = {International Workshop on Approximation and Online Algorithms (WAOA)},
  pages     = {232--251},
  year      = {2019}
}

@inproceedings{ChlamtacMV22,
  author    = {Eden Chlamt{\'a}{\v c} and Yury Makarychev and Ali Vakilian},
  title     = {Approximating Fair Clustering with Cascaded Norm Objectives},
  booktitle = {ACM-SIAM Symposium on Discrete Algorithms (SODA)},
  pages     = {2664--2683},
  year      = {2022}
}

@inproceedings{KleindessnerAM19,
  author    = {Matth{\"a}us Kleindessner and Pranjal Awasthi and Jamie Morgenstern},
  title     = {Fair $k$-Center Clustering for Data Summarization},
  booktitle = {International Conference on Machine Learning (ICML)},
  pages     = {3448--3457},
  year      = {2019}
}

@inproceedings{ChenFLM19,
  author    = {Xingyu Chen and Brandon Fain and Liang Lyu and Kamesh Munagala},
  title     = {Proportionally Fair Clustering},
  booktitle = {International Conference on Machine Learning (ICML)},
  pages     = {1032--1041},
  year      = {2019}
}

@inproceedings{MichaS20,
  author    = {Evi Micha and Nisarg Shah},
  title     = {Proportionally Fair Clustering Revisited},
  booktitle = {International Colloquium on Automata, Languages, and Programming (ICALP)},
  year      = {2020}
}

@inproceedings{MakarychevV21,
  author    = {Yury Makarychev and Ali Vakilian},
  title     = {Approximation Algorithms for Socially Fair Clustering},
  booktitle = {Conference on Learning Theory (COLT)},
  pages     = {3246--3264},
  year      = {2021}
}

@inproceedings{GhadiriSV21,
  author    = {Mehrdad Ghadiri and Samira Samadi and Santosh S. Vempala},
  title     = {Socially Fair $k$-Means Clustering},
  booktitle = {ACM Conference on Fairness, Accountability, and Transparency (FAccT)},
  pages     = {438--448},
  year      = {2021}
}

@inproceedings{MahabadiV20,
  author    = {Sepideh Mahabadi and Ali Vakilian},
  title     = {Individual Fairness for $k$-Clustering},
  booktitle = {International Conference on Machine Learning (ICML)},
  pages     = {6586--6596},
  year      = {2020}
}

@inproceedings{VakilianY22,
  author    = {Ali Vakilian and Mustafa Yal{\c c}{\i}ner},
  title     = {Improved Approximation Algorithms for Individually Fair Clustering},
  booktitle = {International Conference on Artificial Intelligence and Statistics (AISTATS)},
  pages     = {8758--8779},
  year      = {2022}
}

@inproceedings{HotegniMV23,
  author    = {S{\`e}djro Salomon Hotegni and Sepideh Mahabadi and Ali Vakilian},
  title     = {Approximation Algorithms for Fair Range Clustering},
  booktitle = {International Conference on Machine Learning (ICML)},
  pages     = {13270--13284},
  year      = {2023}
}

@inproceedings{JiaSS20,
  author    = {Xinrui Jia and Kshiteej Sheth and Ola Svensson},
  title     = {Fair Colorful $k$-Center Clustering},
  booktitle = {Integer Programming and Combinatorial Optimization (IPCO)},
  pages     = {209--222},
  year      = {2020}
}

@inproceedings{AneggAKZ20,
  author    = {Georg Anegg and Haris Angelidakis and Adam Kurpisz and Rico Zenklusen},
  title     = {A Technique for Obtaining True Approximations for $k$-Center with Covering Constraints},
  booktitle = {Integer Programming and Combinatorial Optimization (IPCO)},
  pages     = {52--65},
  year      = {2020}
}

@inproceedings{EsmaeiliBTD20,
  author    = {Seyed Esmaeili and Brian Brubach and Leonidas Tsepenekas and John Dickerson},
  title     = {Probabilistic Fair Clustering},
  booktitle = {Advances in Neural Information Processing Systems (NeurIPS)},
  year      = {2020}
}

@inproceedings{AbbasiBV21,
  author    = {Mohsen Abbasi and Aditya Bhaskara and Suresh Venkatasubramanian},
  title     = {Fair Clustering via Equitable Group Representations},
  booktitle = {ACM Conference on Fairness, Accountability, and Transparency (FAccT)},
  pages     = {504--514},
  year      = {2021}
}

@inproceedings{ChakrabartyN21,
  author    = {Deeparnab Chakrabarty and Maryam Negahbani},
  title     = {Better Algorithms for Individually Fair $k$-Clustering},
  booktitle = {Advances in Neural Information Processing Systems (NeurIPS)},
  pages     = {13340--13351},
  year      = {2021}
}

@article{NguyenNJ22,
  author  = {Huy L{\^e} Nguyen and Thy Dinh Nguyen and Matthew Jones},
  title   = {Fair Range $k$-Center},
  journal = {arXiv preprint arXiv:2207.11337},
  year    = {2022}
}

@inproceedings{EsmaeiliBSD21,
  author    = {Seyed A. Esmaeili and Brian Brubach and Aravind Srinivasan and John P. Dickerson},
  title     = {Fair Clustering Under a Bounded Cost},
  booktitle = {Advances in Neural Information Processing Systems (NeurIPS)},
  pages     = {14345--14357},
  year      = {2021}
}

@inproceedings{EsmaeiliDDB22,
  author    = {Seyed A. Esmaeili and Sharmila Duppala and John P. Dickerson and Brian Brubach},
  title     = {Fair Labeled Clustering},
  booktitle = {ACM SIGKDD Conference on Knowledge Discovery and Data Mining (KDD)},
  pages     = {327--335},
  year      = {2022},
  doi       = {10.1145/3534678.3539451}
}

@inproceedings{DickersonEMZ23,
  author    = {John P. Dickerson and Seyed A. Esmaeili and Jamie Morgenstern and Claire Jie Zhang},
  title     = {Doubly Constrained Fair Clustering},
  booktitle = {Advances in Neural Information Processing Systems (NeurIPS)},
  year      = {2023}
}

@inproceedings{DickersonEMZ25,
  author    = {John P. Dickerson and Seyed A. Esmaeili and Jamie Morgenstern and Claire Jie Zhang},
  title     = {{SoK}: Fair Clustering: Critique, Caveats, and Future Directions},
  booktitle = {IEEE Conference on Secure and Trustworthy Machine Learning (SaTML)},
  pages     = {698--713},
  year      = {2025},
  doi       = {10.1109/SaTML64287.2025.00044}
}

@article{ChhabraMM21,
  author  = {Anshuman Chhabra and Karina Masalkovait{\.e} and Prasant Mohapatra},
  title   = {An Overview of Fairness in Clustering},
  journal = {IEEE Access},
  volume  = {9},
  pages   = {130698--130720},
  year    = {2021}
}

@inproceedings{DworkHPRZ12,
  author    = {Cynthia Dwork and Moritz Hardt and Toniann Pitassi and Omer Reingold and Richard Zemel},
  title     = {Fairness Through Awareness},
  booktitle = {Innovations in Theoretical Computer Science (ITCS)},
  pages     = {214--226},
  year      = {2012}
}

@inproceedings{Charikar02,
  author    = {Moses Charikar},
  title     = {Similarity Estimation Techniques from Rounding Algorithms},
  booktitle = {ACM Symposium on Theory of Computing (STOC)},
  pages     = {380--388},
  year      = {2002},
  doi       = {10.1145/509907.509965}
}

@inproceedings{IndykT03,
  author    = {Piotr Indyk and Nitin Thaper},
  title     = {Fast Image Retrieval via Embeddings},
  booktitle = {International Workshop on Statistical and Computational Theories of Vision},
  year      = {2003}
}

@misc{UGESP78,
  author       = {{Equal Employment Opportunity Commission} and {Civil Service Commission}
                  and {Department of Labor} and {Department of Justice}},
  title        = {Uniform Guidelines on Employee Selection Procedures},
  howpublished = {29 C.F.R. \S~1607; the ``four-fifths rule'' is \S~1607.4(D)},
  year         = {1978}
}

@inproceedings{FeldmanFMSV15,
  author    = {Michael Feldman and Sorelle A. Friedler and John Moeller and
               Carlos Scheidegger and Suresh Venkatasubramanian},
  title     = {Certifying and Removing Disparate Impact},
  booktitle = {ACM SIGKDD International Conference on Knowledge Discovery and Data Mining (KDD)},
  pages     = {259--268},
  year      = {2015}
}

@article{BarocasS16,
  author  = {Solon Barocas and Andrew D. Selbst},
  title   = {Big Data's Disparate Impact},
  journal = {California Law Review},
  volume  = {104},
  pages   = {671--732},
  year    = {2016}
}

@article{FakcharoenpholRT04,
  author  = {Jittat Fakcharoenphol and Satish Rao and Kunal Talwar},
  title   = {A Tight Bound on Approximating Arbitrary Metrics by Tree Metrics},
  journal = {Journal of Computer and System Sciences},
  volume  = {69},
  number  = {3},
  pages   = {485--497},
  year    = {2004}
}

@inproceedings{Bartal96,
  author    = {Yair Bartal},
  title     = {Probabilistic Approximation of Metric Spaces and its Algorithmic Applications},
  booktitle = {IEEE Symposium on Foundations of Computer Science (FOCS)},
  pages     = {184--193},
  year      = {1996}
}

@inproceedings{Bartal98,
  author    = {Yair Bartal},
  title     = {On Approximating Arbitrary Metrics by Tree Metrics},
  booktitle = {ACM Symposium on Theory of Computing (STOC)},
  pages     = {161--168},
  year      = {1998}
}

@inproceedings{CharikarCGG98,
  author    = {Moses Charikar and Chandra Chekuri and Ashish Goel and Sudipto Guha},
  title     = {Rounding via Trees: Deterministic Approximation Algorithms for Group {S}teiner Trees
               and $k$-Median},
  booktitle = {ACM Symposium on Theory of Computing (STOC)},
  pages     = {114--123},
  year      = {1998}
}

@article{CharikarGTS02,
  author  = {Moses Charikar and Sudipto Guha and {\'E}va Tardos and David B. Shmoys},
  title   = {A Constant-Factor Approximation Algorithm for the $k$-Median Problem},
  journal = {Journal of Computer and System Sciences},
  volume  = {65},
  number  = {1},
  pages   = {129--149},
  year    = {2002}
}

@inproceedings{ByrkaPRST15,
  author    = {Jaros{\l}aw Byrka and Thomas Pensyl and Bartosz Rybicki and Aravind Srinivasan and
               Khoa Trinh},
  title     = {An Improved Approximation for $k$-Median, and Positive Correlation in
               Budgeted Optimization},
  booktitle = {ACM-SIAM Symposium on Discrete Algorithms (SODA)},
  pages     = {737--756},
  year      = {2015}
}

@inproceedings{LiS13,
  author    = {Shi Li and Ola Svensson},
  title     = {Approximating $k$-Median via Pseudo-Approximation},
  booktitle = {ACM Symposium on Theory of Computing (STOC)},
  pages     = {901--910},
  year      = {2013}
}

@inproceedings{CohenAddadGHOS22,
  author    = {Vincent Cohen-Addad and Anupam Gupta and Lunjia Hu and Hoon Oh and David Saulpic},
  title     = {An Improved Local Search Algorithm for $k$-Median},
  booktitle = {ACM-SIAM Symposium on Discrete Algorithms (SODA)},
  pages     = {1556--1612},
  year      = {2022}
}

@inproceedings{GowdaPST23,
  author    = {Kishen N. Gowda and Thomas W. Pensyl and Aravind Srinivasan and Khoa Trinh},
  title     = {Improved Bi-point Rounding Algorithms and a Golden Barrier for $k$-Median},
  booktitle = {ACM-SIAM Symposium on Discrete Algorithms (SODA)},
  pages     = {987--1011},
  year      = {2023}
}

@inproceedings{CohenAddadGLSS25,
  author    = {Vincent Cohen-Addad and Fabrizio Grandoni and Euiwoong Lee and
               Chris Schwiegelshohn and Ola Svensson},
  title     = {A $(2+\varepsilon)$-Approximation Algorithm for Metric $k$-Median},
  booktitle = {ACM Symposium on Theory of Computing (STOC)},
  pages     = {615--624},
  year      = {2025}
}

@inproceedings{CharikarCGGLW25,
  author    = {Moses Charikar and Vincent Cohen-Addad and Ruiquan Gao and
               Fabrizio Grandoni and Euiwoong Lee and Ernest van Wijland},
  title     = {An Improved Greedy Approximation for (Metric) $k$-Means},
  booktitle = {IEEE Symposium on Foundations of Computer Science (FOCS)},
  pages     = {233--240},
  year      = {2025}
}

@article{AryaGKMMP04,
  author  = {Vijay Arya and Naveen Garg and Rohit Khandekar and Adam Meyerson and
             Kamesh Munagala and Vinayaka Pandit},
  title   = {Local Search Heuristics for $k$-Median and Facility Location Problems},
  journal = {SIAM Journal on Computing},
  volume  = {33},
  number  = {3},
  pages   = {544--562},
  year    = {2004}
}

@article{KanungoMNPSW04,
  author  = {Tapas Kanungo and David M. Mount and Nathan S. Netanyahu and Christine D. Piatko and
             Ruth Silverman and Angela Y. Wu},
  title   = {A Local Search Approximation Algorithm for $k$-Means Clustering},
  journal = {Computational Geometry},
  volume  = {28},
  number  = {2-3},
  pages   = {89--112},
  year    = {2004}
}

@article{GuptaT08,
  author  = {Anupam Gupta and Kanat Tangwongsan},
  title   = {Simpler Analyses of Local Search Algorithms for Facility Location},
  journal = {arXiv preprint arXiv:0809.2554},
  year    = {2008}
}

@article{GuhaK99,
  author  = {Sudipto Guha and Samir Khuller},
  title   = {Greedy Strikes Back: Improved Facility Location Algorithms},
  journal = {Journal of Algorithms},
  volume  = {31},
  number  = {1},
  pages   = {228--248},
  year    = {1999}
}

@article{Gonzalez85,
  author  = {Teofilo F. Gonzalez},
  title   = {Clustering to Minimize the Maximum Intercluster Distance},
  journal = {Theoretical Computer Science},
  volume  = {38},
  pages   = {293--306},
  year    = {1985}
}

@article{HochbaumS85,
  author  = {Dorit S. Hochbaum and David B. Shmoys},
  title   = {A Best Possible Heuristic for the $k$-Center Problem},
  journal = {Mathematics of Operations Research},
  volume  = {10},
  number  = {2},
  pages   = {180--184},
  year    = {1985}
}

@inproceedings{CohenAddadSS21,
  author    = {Vincent Cohen-Addad and David Saulpic and Chris Schwiegelshohn},
  title     = {A New Coreset Framework for Clustering},
  booktitle = {ACM Symposium on Theory of Computing (STOC)},
  pages     = {169--182},
  year      = {2021}
}

@article{Li17,
  author  = {Shi Li},
  title   = {On Uniform Capacitated $k$-Median Beyond the Natural {LP} Relaxation},
  journal = {ACM Transactions on Algorithms},
  volume  = {13},
  number  = {2},
  pages   = {1--18},
  year    = {2017}
}

@inproceedings{AdamczykBMMW19,
  author    = {Marek Adamczyk and Jaros{\l}aw Byrka and Jan Marcinkowski and
               Syed Mohammad Meesum and Micha{\l} W{\l}odarczyk},
  title     = {Constant-Factor {FPT} Approximation for Capacitated $k$-Median},
  booktitle = {European Symposium on Algorithms (ESA)},
  pages     = {1:1--1:14},
  year      = {2019}
}

@inproceedings{CohenAddadL19,
  author    = {Vincent Cohen-Addad and Jason Li},
  title     = {On the Fixed-Parameter Tractability of Capacitated Clustering},
  booktitle = {International Colloquium on Automata, Languages, and Programming (ICALP)},
  pages     = {41:1--41:14},
  year      = {2019}
}

@inproceedings{CohenAddad20,
  author    = {Vincent Cohen-Addad},
  title     = {Approximation Schemes for Capacitated Clustering in Doubling Metrics},
  booktitle = {ACM-SIAM Symposium on Discrete Algorithms (SODA)},
  pages     = {2241--2259},
  year      = {2020}
}

@inproceedings{ByrkaFRS15,
  author    = {Jaros{\l}aw Byrka and Krzysztof Fleszar and Bartosz Rybicki and Joachim Spoerhase},
  title     = {Bi-Factor Approximation Algorithms for Hard Capacitated $k$-Median Problems},
  booktitle = {ACM-SIAM Symposium on Discrete Algorithms (SODA)},
  pages     = {722--736},
  year      = {2015}
}

@article{AggarwalPFTKKZ10,
  author  = {Gagan Aggarwal and Rina Panigrahy and Tom{\'a}s Feder and Dilys Thomas and
             Krishnaram Kenthapadi and Samir Khuller and An Zhu},
  title   = {Achieving Anonymity via Clustering},
  journal = {ACM Transactions on Algorithms},
  volume  = {6},
  number  = {3},
  pages   = {49:1--49:19},
  year    = {2010}
}

@book{Schrijver03,
  author    = {Alexander Schrijver},
  title     = {Combinatorial Optimization: Polyhedra and Efficiency},
  publisher = {Springer},
  year      = {2003}
}

@article{anand2026socially,
  title={Socially fair clustering: Parameterized approximation and local search},
  author={Anand, Aditya and Makarychev, Yury and Shan, Liren},
  journal={arXiv preprint arXiv:2608.30026},
  year={2026}
}

@inproceedings{bateni2024scalable,
  title={A scalable algorithm for individually fair k-means clustering},
  author={Bateni, MohammadHossein and Cohen-Addad, Vincent and Epasto, Alessandro and Lattanzi, Silvio},
  booktitle={International Conference on Artificial Intelligence and Statistics},
  pages={3151--3159},
  year={2024},
  organization={PMLR}
}

@article{ebbens2024subquadratic,
  title={A subquadratic time approximation algorithm for individually fair k-center},
  author={Ebbens, Matthijs and Funk, Nicole and H{\"o}ckendorff, Jan and Sohler, Christian and Weil, Vera},
  journal={arXiv preprint arXiv:2412.04943},
  year={2024}
}

\section{Removing the integrality assumption on $t$}
\label{sec:realt}

Every result above assumed $t \in \Z_{\ge 1}$, as does all prior work on exact pairwise
fairness. The assumption can matter elsewhere: \citet{BandyapadhyayCFJ25} show
that for $k$-center the natural relaxation acquires an $\Omega(k)$ integrality gap as
soon as $t$ is fractional, even with two groups. Our algorithm never forms that
relaxation, and the only place integrality was used is
Lemma~\ref{lem:base}(ii). We now recover it with a two-parameter state.

For real $t\ge1$ the counts allowed at a center of base $b$ are $\{b,\dots,\lfloor tb\rfloor\}$. The upper end is no longer proportional to $b$, and Lemma~\ref{lem:base}(ii) fails: for $t=3/2$, two leaves of base $1$ can together hold only $\{1\}+\{1\}=\{2\}$, whereas a single leaf of base $2$ can hold $\{2,3\}$. One integer per subtree is therefore not enough. Two are: we carry the sum of the upper ends along with the sum of the bases.

For integers $0\le b\le c$ let $J(b,c)=\{b,b+1,\dots,c\}$. Given a base vector $h$ with all $h_f\le n$, the \emph{cap} of a leaf $f$ is $c_f=\min\{\lfloor t\,h_f\rfloor,\,n\}$, and for a node $u$ we put $h(u)=\sum_{f\in F_u}h_f$ and $c(u)=\sum_{f\in F_u}c_f$. A load matrix $y'$ is now called \emph{compatible} with $h$ if $y'_{fa}\in J(h_f,c_f)$ for all $f$ and $a$. Capping at $n$ loses nothing, since no load exceeds $n$, and it keeps $c(u)\le mn$.

\begin{lemma}
\label{lem:realbase}
Let $t\ge1$ be real.
\begin{enumerate}
\item[(i)] If $v\in\Z_{\ge0}^{\ell}$ satisfies $\max_av_a\le t\min_av_a$ and $v_a\le n$ for all $a$, then $v_a\in J(b,c)$ for all $a$, where $b=\min_av_a$ and $c=\min\{\lfloor tb\rfloor,n\}$.
\item[(ii)] Conversely, if $v_a\in J(b,c)$ for all $a$, where $c\le\lfloor tb\rfloor$, then $\max_av_a\le t\min_av_a$.
\item[(iii)] Let $b_i\le c_i$ be nonnegative integers for $i=1,\dots,s$. If $x_i\in J(b_i,c_i)$ for every $i$, then $\sum_ix_i\in J\bigl(\sum_ib_i,\sum_ic_i\bigr)$, and every element of $J\bigl(\sum_ib_i,\sum_ic_i\bigr)$ can be written in this way.
\end{enumerate}
\end{lemma}

\begin{proof}
(i) $b\le v_a$ by definition, and $v_a\le\max_av_a\le tb$; since $v_a$ is an integer, $v_a\le\lfloor tb\rfloor$, and $v_a\le n$ by hypothesis.
(ii) $\max_av_a\le c\le\lfloor tb\rfloor\le tb\le t\min_av_a$.
(iii) As for Lemma~\ref{lem:base}(ii): the lower and the upper ends add up, and for the converse one starts from $x_i=b_i$ and repeatedly raises a single $x_i$ with $x_i<c_i$ by one, until the sum is the given element.
\end{proof}

\begin{theorem}[Arbitrary $t$]
\label{thm:realt}
Theorem~\ref{thm:kz}, Theorem~\ref{thm:median} and Corollary~\ref{cor:means} hold for every real $t\ge1$, with the same bounds on the cost. The repair step uses $O(mn^{2})$ states per node and $O(m^{2}n^{4})$ time for combining two tables, instead of $O(n)$ and $O(n^{2})$; it remains polynomial.
\end{theorem}

\begin{proof}
Only Section~\ref{sec:repair} changes, and only in that the window $I(h(u))$ of a node is replaced everywhere by $J(h(u),c(u))$. Thus $\delta^a_u(b,c)=\dist\bigl(y_a(u),J(b,c)\bigr)$, $\delta_u(b,c)=\sum_a\delta^a_u(b,c)$, the surrogate is $\Phi_T(h)=\sum_uw_u\,\delta_u\bigl(h(u),c(u)\bigr)$, and $h$ is admissible if $N_a\in J(h(\rt),c(\rt))$ for all $a$, that is, if $h(\rt)\le\min_aN_a$ and $c(\rt)\ge\max_aN_a$.

\emph{Pass 1.} The table of a node is indexed by the pairs $(b,c)$ with $0\le b\le n$ and $b\le c\le mn$, and $V_u(b,c)$ is the best value of the subtree of $u$ over the choices of leaf bases with $h(u)=b$ and $c(u)=c$. A leaf $f$ admits only the pairs $(b,\min\{\lfloor tb\rfloor,n\})$, with value $w_f\,\delta_f(b,c)$. At an internal node, \eqref{eq:dpup} becomes $V_u(b,c)=w_u\,\delta_u(b,c)+\min\sum_iV_{u_i}(b_i,c_i)$, the minimum being over all ways of writing $(b,c)=\sum_i(b_i,c_i)$. The second coordinate must be allowed to reach $mn$: it is a sum of up to $m$ leaf caps, each at most $n$, and it can exceed $n$ even though no load does. Every admissible base vector is represented in the tables, because it has $h(u)\le\min_aN_a\le n$ and $c(u)\le mn$ at every node. At the root we minimize over the admissible pairs. Such pairs exist: giving one leaf the base $\min_aN_a$ and all other leaves the base $0$ yields $b=\min_aN_a$ and $c=\min\{\lfloor t\min_aN_a\rfloor,n\}\ge\max_aN_a$, because $\max_aN_a$ is an integer that is at most $t\min_aN_a$ and at most $n$.

\emph{Pass 2} is unchanged, with $q_i$ the element of $J(h(u_i),c(u_i))$ closest to $y_a(u_i)$ and with $c(u_i)$ in place of $t\,h(u_i)$ as the upper limit of the corrections. They succeed because $\sum_ih(u_i)=h(u)\le x_a(u)\le c(u)=\sum_ic(u_i)$.

\emph{Analysis.} The three uses of Lemma~\ref{lem:base} become uses of Lemma~\ref{lem:realbase}. In Lemma~\ref{lem:lower}, part~(iii) gives $y'_a(u)\in J(h(u),c(u))$ for every compatible $y'$, and the rest of the proof is unchanged. In Lemma~\ref{lem:recon}, the inequality \eqref{eq:reconstruct} and the summation that follows it do not involve $t$ at all, and the output satisfies $\bar y_{fa}\in J(h_f,c_f)$ with $c_f\le\lfloor t\,h_f\rfloor$, so its rows are fair by part~(ii). In Lemma~\ref{lem:repair}, a fair comparison matrix $y^\star$ has all entries at most $n$, so by part~(i) it is compatible with the vector $h^\star$ of its row minima, and the chain of inequalities goes through.

\emph{Running time.} A table has $(n+1)(mn+1)=O(mn^{2})$ entries, and combining two tables takes time quadratic in that, $O(m^{2}n^{4})$. (Since no count $y_a(u)$ exceeds $n$, the forced traffic depends on $c$ only through $\min\{c,n\}$. One may therefore store $\min\{c(u),n\}$ instead and add caps with saturation at $n$, which brings the tables back to $O(n^{2})$ entries. We do not need this refinement.)
\end{proof}

\section{Lower bounds}
\label{sec:limits}

\subsection{$\Omega(\ell t)$ integrality gap of the natural relaxation}

Prior polynomial-time work on exact pairwise fairness rounds the following relaxation.
Given candidate centers $F$, use variables $x_{pf} \ge 0$ indicating that $p$ is
assigned to $f$:
\begin{equation}
\label{eq:lp}
\begin{aligned}
  \min \quad & \textstyle\sum_{p \in P}\sum_{f \in F} d(p,f)^{z}\, x_{pf} \\
  \text{s.t.}\quad
   & \textstyle\sum_{f \in F} x_{pf} = 1 && \forall p \in P,\\
   & \textstyle\sum_{p \in P_a} x_{pf} \;\le\; t \sum_{p \in P_b} x_{pf}
     && \forall f \in F,\ \forall a,b \in [\ell],\\
   & x_{pf} \ge 0 .
\end{aligned}
\end{equation}

\begin{proposition}
\label{prop:lpgap}
For every $\ell \ge 2$, every real $t \ge 2$ and every $z \ge 1$, relaxation
\eqref{eq:lp} has integrality gap at least $(\ell-1)t$. The witnessing instance has two
candidate centers at distance $1$ on a line.
\end{proposition}

\begin{proof}
Place two locations $u,v$ with $d(u,v) = 1$ and $F = \{u,v\}$. At $u$ put one point of
each group $1,\dots,\ell$. At $v$ put one point of each group $1,\dots,\ell-1$ and none
of group $\ell$. Then $N_a = 2$ for $a < \ell$ and $N_\ell = 1$, and the instance is
feasible since $2 \le t \cdot 1$.

\smallskip
\emph{Integral optimal solution is $\ell-1$.} A nonempty fair cluster must contain a point of
group $\ell$, and there is only one such point, so at most one cluster is nonempty and
all points go to one location. Sending everything to $u$ moves the $\ell-1$ points at
$v$, at cost $\ell-1$; the resulting counts are $(2,\dots,2,1)$, which is fair because
$2 \le t$. Sending everything to $v$ costs $\ell$. Hence the integral optimal is exactly
$\ell-1$.

\smallskip
\emph{Fractional optimal solution is at most $1/t$.} Keep every point where it is, except assign
the group-$\ell$ point to $v$ to extent $1/t$ and to $u$ to extent $1 - 1/t$. At $v$ the
group loads are $(1,\dots,1,1/t)$, and the constraint requires $1 \le t\cdot(1/t) = 1$,
which holds. At $u$ they are $(1,\dots,1,1-1/t)$, and the constraint requires
$1 \le t(1-1/t) = t-1$, which holds because $t \ge 2$. The objective is
$(1/t)\,d(u,v)^{z} = 1/t$.
\end{proof}

Since $\ell$ can be as large as $n$ and $t$ as large as $n$, no rounding of
\eqref{eq:lp} alone can produce a guarantee that is independent of $\ell$ and $t$, which is why the $O(k^{2}\ell t)$ bound of \citet{BandyapadhyayCFJMV26} carries those parameters.

\subsection{Tightness of $O(\log k)$ for $k$-median}
\label{sec:tightness}

\begin{proposition}
\label{prop:conditional}
If pairwise fair $k$-median admits a polynomial-time $o(\log k)$-approximation, even
for $\ell = 2$, then soft uniform capacitated $k$-median admits a
polynomial-time $o(\log k)$-approximation.
\end{proposition}

\begin{proof}
\citet{BandyapadhyayCFJMV26} show that an $\alpha$-approximation for pairwise fair
$k$-median with $\ell=2$ yields a $(1+\eps)\alpha$-approximation for soft uniform
capacitated $k$-median, for every constant $\eps > 0$, and the reduction preserves $k$.
Composing with $\alpha = o(\log k)$ gives the claim.
\end{proof}

For soft uniform capacitated $k$-median the best known polynomial-time guarantee that
violates neither the capacities nor the number of centers is $O(\log k)$, obtained by
probabilistic tree embedding followed by exact dynamic programming on the tree
\citep{CharikarCGG98,Bartal96}, and whether a constant factor is possible is a
well-known open problem \citep{Li17,CohenAddadL19,CohenAddad20}. Theorem~\ref{thm:median}
therefore matches that barrier exactly, and no improvement is possible without progress on capacitated $k$-median, for which, despite extensive research, the best known polynomial-time approximation factor remains $O(\log k)$.

\end{document}